\documentclass[journal]{IEEEtran}
	\usepackage{caption}
	\usepackage{graphicx}
	\usepackage{cite}
	\usepackage{amsmath}
	\usepackage{mathrsfs}
	\usepackage{amssymb}
	\usepackage{algorithm}
	\usepackage{algorithmic}
	\usepackage{setspace}
	\usepackage{booktabs}
	\usepackage{placeins}
	\usepackage{subcaption}
	\usepackage{lipsum}
	\makeatletter
	\renewcommand{\maketag@@@}[1]{\hbox{\m@th\normalsize\normalfont#1}}
	\makeatother
\begin{document}

		\title{Integrated Communication and Remote Sensing in LEO Satellite Systems: Protocol, Architecture and Prototype}
	
		\author{\IEEEauthorblockN{
			\normalsize Yichao Xu, Xiaoming Chen, Ming Ying, and Zhaoyang Zhang}
			\thanks{Yichao Xu, Xiaoming Chen, Ming Ying and Zhaoyang Zhang are with the College of Information	Science and Electronic Engineering, Zhejiang University, Hangzhou, 310027,	China (e-mail: \{yichao\_xu, chen\_xiaoming, ming\_ying, ning\_ming\}@zju.edu.cn). }
		}\maketitle

	\begin{abstract}
		In this paper, we explore the integration of communication and synthetic aperture radar (SAR)-based remote sensing in low Earth orbit (LEO) satellite systems to provide real-time SAR imaging and information transmission.
		Considering the high-mobility characteristics of satellite channels and limited processing capabilities of satellite payloads, we propose an integrated communication and remote sensing architecture based on an orthogonal delay-Doppler division multiplexing (ODDM) signal waveform.
		Both communication and SAR imaging functionalities are achieved with an integrated transceiver onboard the LEO satellite, utilizing the same waveform and radio frequency (RF) front-end.
		Based on such an architecture, we propose a transmission protocol compatible with the 5G NR standard using downlink pilots for joint channel estimation and SAR imaging.
		Furthermore, we design a unified signal processing framework for the integrated satellite receiver to simultaneously achieve high-performance channel sensing, low-complexity channel equalization and interference-free SAR imaging.
		Finally, the performance of the proposed integrated system is demonstrated through comprehensive analysis and extensive simulations in the sub-6 GHz band.
		Moreover, a software-defined radio (SDR) prototype is presented to validate its effectiveness for real-time SAR imaging and information transmission in satellite direct-connect user equipment (UE) scenarios within the millimeter-wave (mmWave) band.
	\end{abstract}
	
	
	\section{Introduction}
	\label{sec1}
	With the increasing demand for timely remote sensing in applications such as emergency rescue and weather monitoring, low Earth orbit (LEO) remote sensing satellite systems are continually evolving.
	However, existing standalone remote sensing systems suffer from significant response latency, primarily due to the limited visibility and transmission capacity of remote sensing satellites, and sometimes even requiring assistance from communication satellites \cite{cite1}.
	Concurrently, LEO broadband communication satellite systems, known for their low latency and minimal propagation loss, have emerged as a focal point in the development of next-generation non-terrestrial networks, and have already begun to achieve commercialization as exemplified by SpaceX's Starlink \cite{cite2}.
	To this end, the integration of communication and remote sensing in LEO satellite systems has naturally evolved, and is expected to enable real-time remote sensing and transmission in satellite direct-connect user equipment (UE) scenarios.
	However, due to their historical separation, communication and remote sensing satellite systems exhibit a range of differences, including signal representation, transmission protocols, and signal processing frameworks \cite{cite3}.
	
	Considering similar channel characteristics between satellite communication and synthetic aperture radar (SAR) remote sensing systems, the LEO satellite can perform shared signal transmission with a unified waveform, and thus realizes joint communication and SAR imaging by extracting the radar-related parameters from received signals.
	In this context, the waveform design plays a crucial role in seamlessly integrating both communication and SAR functionalities on the same satellite platform.
	Several studies have been conducted to integrate communication and radar sensing signals by leveraging the degrees of freedom of resources across time \cite{cite4}, frequency \cite{cite5}, spatial \cite{cite6}, and other domains.
	Generally, these methods organize discrete functional signals into a unified waveform, avoiding interference between communication and radar sensing through orthogonal resource allocation \cite{cite7}, or mitigating interference between communication and radar sensing by optimizing non-orthogonal transmission \cite{OptMIMO}.
	However, fulfilling both performance requirements of communication and remote sensing while adhering to limited resources remains a challenging task.
	
	To this end, advanced waveforms have been explored for dual-functional signal sharing in integrated sensing and communication (ISAC) systems. Specifically, orthogonal frequency division multiplexing (OFDM) signals were employed in \cite{cite8} to develop jointly correlated pulses, enabling high data rates and precise target detection.
	Moreover, an OFDM-LFM scheme was proposed to use linear frequency modulated (LFM) signals as subcarriers within the OFDM modulation, mitigating the vulnerability of OFDM signals to frequency shifts \cite{OFDM-LFM}.
	However, inter-carrier interference (ICI) and inter-symbol interference (ISI) caused by doubly dispersive channels of LEO satellite remain challenging issues when using OFDM's time-frequency (TF)-domain modulation.
	To address these, Hadani et al. introduced a two-dimensional orthogonal time frequency space (OTFS) modulation, where information symbols are placed in the delay-Doppler (DD) domains \cite{cite9}.
	Notably, the delay and Doppler parameters in OTFS closely resemble those used in radar systems, which has prompted  several preliminary studies to explore the potential application of OTFS in ISAC systems \cite{cite10}.
	Since channel parameters are crucial for both OTFS-based communication and radar sensing, effectively performing channel sensing becomes a key issue in OTFS-ISAC systems \cite{cite11}.
	A series of correlation-based methods have been applied to DD-domain channel estimation using pseudo-noise (PN) sequences or cyclic prefix (CP)-based OTFS signals \cite{cite12}, \cite{cite13}, \cite{cite14}.
	Raviteja et al. proposed a low-complexity single-pilot-aided (SPA) method \cite{cite15}, improving the efficiency of both channel estimation and data detection.
	The inherent separability and sparsity of DD-domain channels have also motivated researchers to employ compressed sensing (CS) techniques to tackle the estimation challenges \cite{cite17}.
	Furthermore, Mishra et al. designed a superimposed pilot (SuP)-based channel estimation and data detection framework for OTFS systems to achieve higher spectral efficiency \cite{mishra2021otfs}.
	
	However, existing research on channel sensing of OTFS mainly focuses on communication-only or near-field localization functions \cite{cite18}, \cite{cite19}, \cite{cite20}, with a lack of further studies and designs specifically tailored to satellite remote sensing scenarios.
	Hence, LEO satellite-based integration still presents several challenges, primarily attributed to the high-mobility LEO satellite, as well as the unique requirements of remote sensing payloads, particularly SAR.
	Currently, SAR has become one of the commonly used methods in satellite remote sensing due to its high-resolution imaging and all-weather observation capabilities.
	SAR systems periodically transmit probe signals with a certain pulse repetition frequency (PRF) restriction, and the echo signals from scatters in different range cells is overlapped.
	To mitigate this interference, conventional SAR systems utilize LFM pulses with a large time-bandwidth product and perform matched filtering (MF)-based pulse compression to achieve SAR imaging.
	However, matched filtering inevitably introduces inter-range-cell interference (IRCI) due to the side lobes in the ambiguity function of probe signals.
	Since the echo signals from different range cells can be viewed as received signals from paths with varying delays in multi-path communication, a CP-based OFDM-SAR technology was proposed to achieve high resolution IRCI-free SAR imaging \cite{cite21}, \cite{cite22}.
	However, these methods are not directly applicable to spaceborne systems due to the Doppler sensitivity of OFDM waveforms.
	To this end, Sharma et al. employ OTFS waveform to perform low-complexity SAR imaging in the presence of intra-pulse Doppler shift \cite{cite23}.
	However, this approach accounts for the Doppler shift known in advance, necessitating further research for practical applicability.
	Since existing research predominantly focuses on designing SAR imaging methods based on novel probe signals, the integration of communication and SAR functionalities still lacks specific transmission protocols and operational architectures for support.
	Thus, this work aims to propose practical and feasible architecture, protocol, and algorithm for such an integrated communication and remote sensing system.
	
	The recently proposed orthogonal delay-Doppler division multiplexing (ODDM) waveform has garnered increasing interest due to its adaptability to doubly dispersive channels with the delay-Doppler domain orthogonal pulse (DDOP) \cite{cite24}.
	Although adopting the same DD-domain modulation as OTFS, ODDM employs the DDOP to achieve an exact DD-domain input-output (IO) relationship, enabling precise channel estimation and signal detection, whereas OTFS just an approximation due to the complicated TF-domain ISI and ICI.
	Moreover, the authors in \cite{cite24} demonstrated that the ODDM outperforms the traditional OTFS in terms of out-of-band emission and bit error rate (BER).
	To this end, ODDM modulation exhibits strong potential as an effective solution for integrated communication and SAR remote sensing in LEO satellite systems.

	In this paper, we aim to perform integrated communication and remote sensing in LEO satellite systems to achieve real-time SAR imaging and information transmission.
	It is particularly challenging due to the large Doppler shift and multipath fading effects in satellite channels and the unique signal transmission and processing demands for SAR imaging.
	To address these challenges, we first propose an integrated communication and remote sensing system architecture with the ODDM signal waveform, which guarantees robustness to fast time-varying channels.
	To be compatible with terrestrial 5G systems, we adopt a wireless frame configuration of 5G NR standard, replacing the underlying waveform with ODDM signals and rearranging the patterned symbols of each function in the DD domain.
	Based on the advanced ISAC waveform, we further propose unified signal processing framework and algorithm in satellite receiver to achieve low-complexity ODDM communication reception and IRCI-free SAR imaging. Moreover, we validate the feasibility of the proposed integrated communication and remote sensing system using a software-defined radio (SDR)-based prototype platform in millimeter-wave (mmWave) band.
	The contributions of the paper are summarized as follows:
	\begin{itemize}
		\item [1)]
		We propose a system architecture for integrated communication and remote sensing based on unified ODDM waveform and RF front-end.
		\item [2)]
		We design a wireless frame structure compatible with 5G NR standard, which is flexibly adjustable to the PRF restriction of SAR imaging.
		\item [3)]
		We give an effective integrated communication and remote sensing algorithm for satellite receiver, including channel sensing, communication equalization and SAR imaging.
		\item [4)]
		We construct a SDR-based prototype system, and validate the feasibility of real-time SAR imaging and information transmission.
	\end{itemize}
	
	The paper is organized in the following manner.
	In Section \ref{sec2}, we present the system model and introduce the work mode of the integrated communication and remote sensing system.
	Section \ref{sec3} proposes a transmission protocol for integrated communication and remote sensing.
	Then, we specifically design a transceiver architecture, proposing an integrated communication and remote sensing algorithm in Section \ref{sec4}.
	Section \ref{sec5} provides simulation results and conducts prototype verification with a SDR-based platform.
	Finally, Section VI concludes the paper.
	
	\textit{Notation:} We use bold upper letters and bold lower letters to represent matrices and column vectors, respectively. $(\cdot)^\mathrm{T}$, $(\cdot)^\mathrm{-1}$ and $(\cdot)^\mathrm{H}$ indicate transpose, Hermitian transpose and inverse, respectively. $\mathrm{diag}(\cdot)$ and $\mathrm{diag}^{-1}_{M,N}(\cdot)$ give transformation and inversion of vector to square diagonal matrix. $\mathbf{I}_M$ is $M\times M$ identity matrix, $\mathbf{0}$ denotes an all-zero matrix, and $\mathbf{F}_{N}$ represents $N\times N$ normalized $N$-point discrete Fourier transform matrix.
	We use $\lVert \cdot \rVert$ to denote the $L_2$-norm of a vector or matrix, and $\vert \cdot \vert$ to denote absolute value.
	$(\cdot)_N$ and $\oslash$ represent modulo-N and point-wise division, respectively.
	$\mathbb{E}$ denotes expectation, and $\mathbb{C}^{M\times N}$ denotes the set of $M\times N$ dimensional complex matrix. $\mathcal{N}(\mu,\sigma^{2})$ is the Gaussian distribution with mean $\mu$ and variance $\sigma^2$.
	
	\section{System Model}
	\label{sec2}
\begin{figure}[tbph!]
		\centering
		\includegraphics[width=1\linewidth]{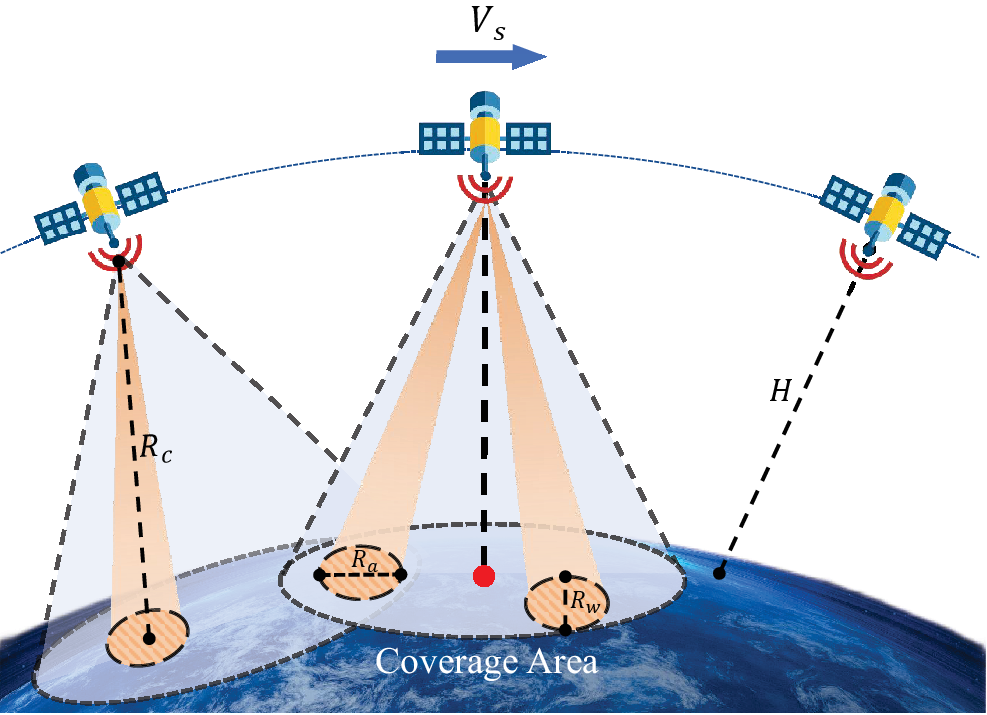}
		\captionsetup{labelformat=empty}
		\caption{\small Fig.~1.~~Model of multi-beam LEO satellite constellation.}
		\label{fig1}
	\end{figure}

	We consider an integrated communication and SAR remote sensing system on a LEO satellite equipped with multi-beam phased array antennas. As shown in Fig. \ref{fig1}, the LEO satellite is part of a Walker Delta configuration-based constellation\footnote{In the Walker Delta constellation, all satellites maintain the same orbit altitude and are evenly distributed on circular orbits with a common inclination angle. Compared to other configurations, such as the Walker Star \cite{walker1970circular} and Rosette \cite{ballard2007rosette} constellations, the Walker Delta configuration offers advantages in terms of unified global coverage, efficient satellite utilization, minimized inter-satellite interference, scalability, and ease of design and implementation. This configuration has been extensively studied and validated, making it widely applicable in satellite communications and remote sensing systems \cite{walker1984satellite}.}.
	Although the coverage areas of adjacent satellites may overlap, different beams serve distinct ground cells, and the inter-beam interference can be effectively mitigated through multi-color reuse \cite{8467316} or multi-beam precoding \cite{devillers2011joint} techniques. As a result, each LEO satellite can be regarded as independently transmitting shared signals to its beam-covered area, characterized by a swath width of $R_w$, an aperture length of $R_a$ and a beam center slant range of $R_c$. Within the scope of joint communication-SAR beam, the integrated system is capable of concurrently supporting satellite-to-ground communication with multiple terrestrial user devices and performing SAR imaging by exploiting the echo signals.
	
\begin{figure}[tbph!]
	\centering
	\includegraphics[width=1\linewidth]{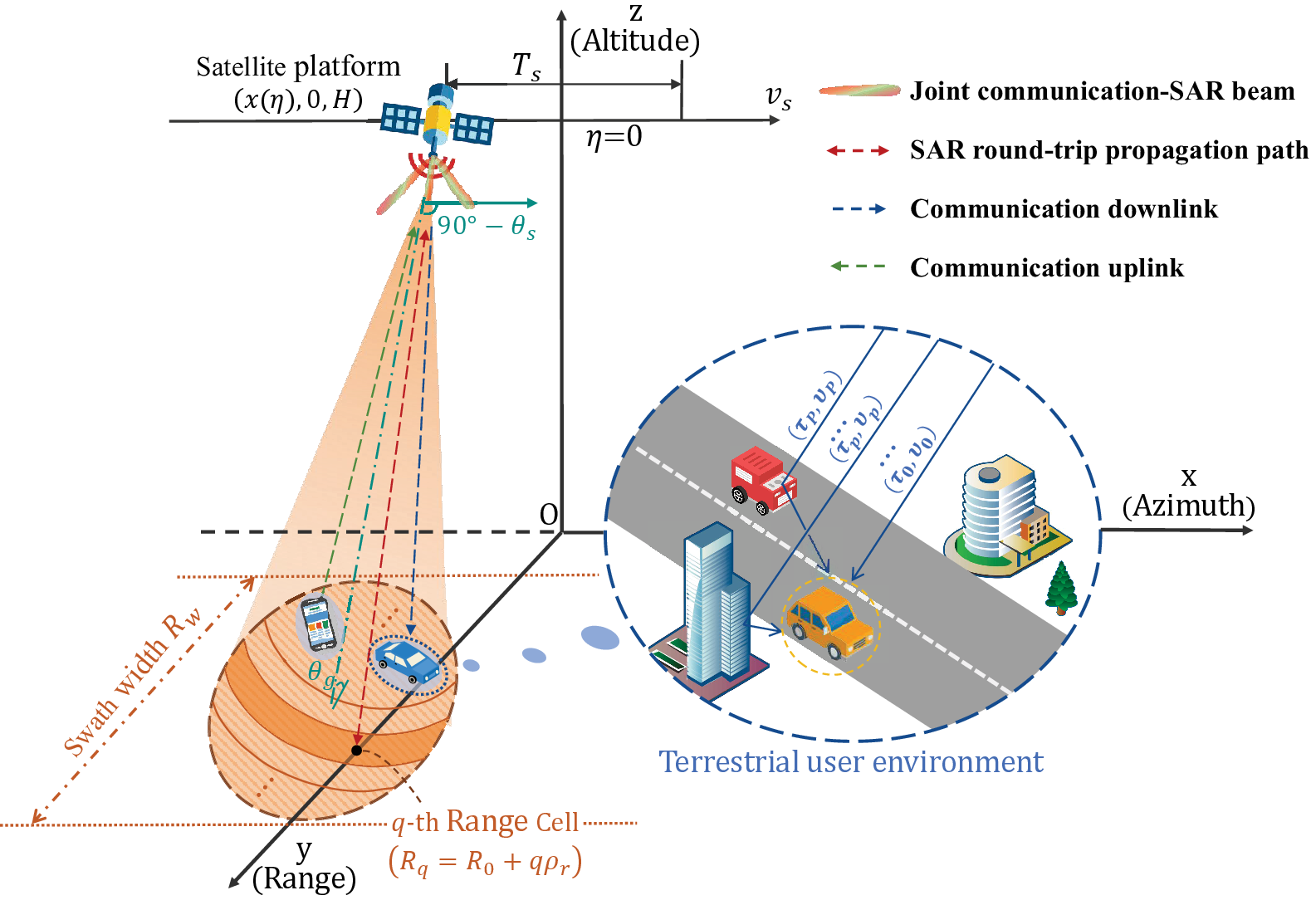}
	\captionsetup{labelformat=empty}
	\caption{\small Fig.~2.~~Model of integrated communication and remote sensing in LEO satellite systems.}
	\label{fig2}
	\end{figure}

	Specifically, the integrated system employs monostatic broadside stripmap SAR geometry with a squint angle ${\theta }_s$ and a grazing angle ${\theta }_g$, as depicted in Fig. \ref{fig2}.
	The satellite platform is moving parallel to the $x$-axis at a constant velocity $V_s$, with an instantaneous coordinate given by  $\left(x(\eta) ,0,H\right)$.
	Here, $x(\eta)$ denotes the coordinate position of the satellite along the direction of motion, given as $x(\eta)=V_s\eta$, where $\eta$ represents the relative azimuth time referenced to the beam center crossing time, typically defined over the synthetic aperture duration $T_s$, and $H$ denotes the orbit altitude of the satellite platform.
	
	In what follows, we first specify the channel models for communication and SAR remote sensing, respectively.
	Subsequently, we introduce an ODDM waveform adaptive to the LEO satellite channel conditions.
	Finally, we discuss the work mode of the integrated system.

	\subsection{Channel Model}
	\subsubsection{Communication Channel Model}
	Due to the high mobility of LEO satellites and the complex multipath fading effects during transmission, simultaneous time and frequency dispersion occurs in the satellite-to-ground channel. According to the signal propagation characteristics of the LEO satellite communication \cite{cite25}, such a doubly dispersive channel can be effectively represented in a DD spreading function.
	Without loss of generality, the spreading function from the LEO satellite to any terrestrial user device can be expressed as
	\begin{equation} \label{eq1}
		h^\mathrm{c}(\tau, \upsilon) = g\left( \sqrt{\frac{K_r}{K_r + 1}} h^{\scriptscriptstyle \mathrm{LOS}}(\tau, \upsilon)
		+ \sqrt{\frac{1}{K_r + 1}} h^{\scriptscriptstyle \mathrm{NLOS}}(\tau, \upsilon) \right),
	\end{equation}
	where $\tau$ and $\upsilon$ refer to the delay and Doppler variables of the DD-domain channel, respectively, and $g$ represents the channel's large-scale fading, given by
	\begin{equation} \label{eq2}
		g\mathrm{=}\sqrt{\mathrm{(}\frac{c}{\mathrm{4}\pi f_0d_0}{\mathrm{)}}^{\mathrm{2}}\cdot G\omega \cdot \frac{\mathrm{1}}{r}}\mathrm{\ ,}
	\end{equation}
	where $\mathrm{(}\frac{c}{\mathrm{4}\pi f_0d_0}{\mathrm{)}}^{\mathrm{2}}$ denotes the free space loss with $c$ being the light speed, $d_0$ being the propagation distance, $f_0$ being the carrier frequency. $\omega $ represents the satellite antenna gain, $G$ denotes the antenna gain of the user device, and the rain attenuation coefficient $r$ reflects the impact of atmospheric conditions, whose power gain in dB $r^\mathrm{dB}\mathrm{=20}\log_{\mathrm{10}} r$, follows log-normal random distribution $\ln\mathrm{(}r^\mathrm{dB}\mathrm{)}\mathrm{\sim }\mathcal{N}\mathrm{(}{\mu }_r,{\sigma }^{\mathrm{2}}_r\mathrm{)}$.
	
	For channel small-scale fading $\sqrt{\frac{K_r}{K_r + 1}} h^{\scriptscriptstyle \mathrm{LOS}}(\tau, \upsilon)
	+ \sqrt{\frac{1}{K_r + 1}} h^{\scriptscriptstyle \mathrm{NLOS}}(\tau, \upsilon)$, $K_r$ is the Rician factor, $h^{\scriptscriptstyle \mathrm{LOS}}(\tau, \upsilon)$ and $h^{\scriptscriptstyle \mathrm{NLOS}}(\tau, \upsilon)$ denote the line-of-sight (LOS) and non-line-of-sight (NLOS) components of the satellite-to-ground channel, respectively, which can be expressed as
	\begin{subequations} \label{eq3}
		\begin{align}
			h^{\scriptscriptstyle \mathrm{LOS}}(\tau, \upsilon)  &= h_0 \delta(\tau - \tau_0) \delta(\upsilon - \upsilon_0) \label{eq3a} \\
			h^{\scriptscriptstyle \mathrm{NLOS}}(\tau, \upsilon)  &= \sqrt{\frac{1}{P}} \sum_{p=1}^{P} h_p \delta(\tau - \tau_p) \delta(\upsilon - \upsilon_p), \label{eq3b}
		\end{align}
	\end{subequations}
	where $\delta(\cdot)$ is the Dirac delta function, $P$ denotes the number of NLOS paths between the satellite and the user device, ${\tau }_p$, ${\upsilon }_p$ and $h_p$ are the propagation delay, the Doppler shift and the complex-valued gain of the $p$-th path associated with the user device, respectively.
	Notice that the LOS path gain $h_0$ has normalized power and the NLOS path gain satisfies a constraint of $\frac{1}{P}\sum^P_{p=1}\mathbb{E}\{\vert h_p{\vert}^{\mathrm{2}}\}\mathrm{=1}$.
	
	It is widely known that LEO satellite has high mobility and long propagation distances, resulting in large Doppler shifts and the long transmission delays.
	For Doppler shift ${\upsilon }_p$, it can be represented as a sum of two parts, i.e., ${\upsilon }_p\mathrm{=}{{\upsilon }_{\scriptscriptstyle \mathrm{Sat}}+\upsilon }_{\scriptscriptstyle \mathrm{Dev}}\mathrm{cos(}{\mathrm{\Phi }}_p\mathrm{)}$,~$p\in \{0,\dots, P\}$.
	Here, ${\upsilon }_{\scriptscriptstyle \mathrm{Sat}}$ denotes the dominant Doppler shift caused by LEO satellite, which is approximately equal for all paths due to the high altitude of the LEO satellite orbit, i.e., ${\upsilon }_{\scriptscriptstyle \mathrm{Sat}}=\frac{V_s\mathrm{sin}\mathrm{(}{\theta }_s\mathrm{)}}{c}f_0$.
	The Doppler shift caused by user device movement is typically smaller, and varies with the angle of arrival ${\mathrm{\Phi }}_p$ between the direction of device movement and the $p$-th propagation path.
	For propagation delay ${\tau }_p$, it is reasonably assumed that the LOS path has the shortest propagation distance, and the NLOS paths are arranged in ascending order of delay, i.e., ${\tau }_{\mathrm{min}}\mathrm{=}{\tau }_0$ and ${\tau }_{\mathrm{max}}\mathrm{=}{\tau }_P$.
	Thereby, the communication channel model can be simplified as
	\begin{equation} \label{eq4}
		h^\mathrm{c}\mathrm{(}\tau ,\upsilon \mathrm{)=}\sum^P_{p\mathrm{=0}}{\alpha }_p\delta \mathrm{(}\tau -\tau_p\mathrm{)}\delta \mathrm{(}\upsilon -\upsilon_p\mathrm{),}
	\end{equation}
	where ${\alpha }_p$ represents the general complex power gain of the $p$-th path to the user device, given by
	\begin{equation}\label{eq5}
		\alpha_{p} = \begin{cases}
			g\cdot \sqrt{\frac{K_r}{{K_r + 1}}} \cdot h_{0}, & p = 0 \vspace{10pt}\\
			g\cdot \sqrt{\frac{1}{{K_r + 1}}} \sqrt{\frac{1}{{P}}} \cdot h_{p}, &  p = 1, \ldots, {P}
		\end{cases}.
	\end{equation}

	In this way, we provide a channel model for LEO satellite communications, with relevant channel parameters depending on the particular scenario.
	
	\subsubsection{SAR Channel Model}
	In SAR remote sensing, probe signals are periodically transmitted by the satellite platform over relative azimuth time $\eta \in \left[-\frac{T_s}{2}, \frac{T_s}{2}\right]$ at a specified PRF.
	Each probe signal propagates through a round-trip channel, with radar echoes from different azimuth time stored in parallel, forming a two-dimensional raw radar data structure for imaging processing.
	Here, we consider a single set of range-time data at a given azimuth time for SAR channel analysis.
	Let the swath width of the joint beam be $R_w$, the grazing angle be $\theta_g$, and the range resolution be represented as ${\rho }_r\mathrm{=}\frac{c}{\mathrm{2}B_w}$, where $B_w$ is the bandwidth of the probe signal.
	Then, the range profile can be divided into $Q\mathrm{=}\frac{R_w}{{\rho }_r}\mathrm{cos(}{\theta }_g\mathrm{)}$ range cells as depicted in Fig. \ref{fig2}, which correspond to $Q$ paths in the multipath communication model \cite{cite21}.
	Moreover, $R_q$, as the instantaneous slant range between the satellite platform and the $q$-th range cell, is given by $R_q = R_0 + q {\rho}_r$, where $R_0$ is the instantaneous slant range between the satellite platform and the first range cell.
	Notably, the beam center slant range $R_{\mathrm{c}}$ can be determined as $R_{\mathrm{c}}=\frac{1}{Q}\sum_{q=0}^{Q-1}R_q=\frac{H}{\sin({\theta}_g)}$.
	Meanwhile, the mobility velocity $V_s$ of the LEO satellite and the squint angle ${\theta }_s$ of the joint beam will inevitably cause an intra-pulse Doppler shift.
	Since the main scatters of each range cell can be seen as stationary and the size of beam swath is typically much less than the distance between satellite platform and beam centre, the intra-pulse Doppler shift can be considered as same for all range cells.
	Therefore, the SAR remote sensing channel model can be expressed as below
	\begin{equation} \label{eq6}
		h^\mathrm{r}\mathrm{(}\tau ,\upsilon \mathrm{)=}\sum^{Q-\mathrm{1}}_{q\mathrm{=0}}{}\beta_q\delta \mathrm{(}\tau -\tilde{{\tau }}_q\mathrm{)}\delta \mathrm{(}\upsilon -\tilde{\upsilon }_r\mathrm{)},
	\end{equation}
	where $\tilde{{\tau }}_q$ is the round-trip propagation delay of the $q$-th range cell, namely $\tilde{{\tau }}_q\mathrm{=}\frac{\mathrm{2}R_q}{c}=\frac{\mathrm{2}R_0}{c}+\frac{q}{B_w}$, and $\tilde{{\upsilon }}_r$
	denotes the common intra-pulse Doppler shift, given as $\tilde{\upsilon }_r\mathrm{=}\frac{2V_s\mathrm{sin}\mathrm{(}{\theta }_s\mathrm{)}}{c}f_0$.
	$\beta_q$ represents the complex power gain associated with the $q$-th range cell according to basic radar equation as \cite{cite26}
	\begin{equation} \label{eq7}
		\beta_q\mathrm{=}\sqrt{G_q \cdot \frac{c^2}{(4\pi)^3 f_0^2R_q^4}\cdot {\omega }^2\cdot \frac{\mathrm{1}}{r}} ,
	\end{equation}
	where $G_q$ is the radar cross section (RCS) coefficient caused by the scatters in the $q$-th range cell. Notably, in SAR imaging, range reconstruction refers to extract RCS coefficients of each range cell within the beam swath, which can equivalently be realized by estimating the path gains $\{\beta_q\}_{q=0}^{Q-1}$ at each azimuth time $\eta$.
	
	For both the communication and SAR remote sensing channels, the IO relation between the transmitted and received baseband signals can be efficiently modeled with a generalized DD spreading function, $h^\star(\tau ,\upsilon )$, as
	\begin{equation} \label{eq8}
		r^\star\mathrm{(}t\mathrm{)}=\iint h^\star\mathrm{(}\tau, \upsilon\mathrm{)}s\mathrm{(}t-\tau\mathrm{)}e^{j2\pi\upsilon\mathrm{(}t-\tau\mathrm{)}}e^{-j2\pi f_0\tau}d\tau d\upsilon,
	\end{equation}
	where the superscript $\star$ is a placeholder for the channel type: $\star=\mathrm{c}$ refers to the communication channel, while $\star=\mathrm{r}$ denotes to the SAR remote sensing channel. $s(t)$ represents the baseband signals transmitted from the LEO satellite and $r^\star(t)$ denotes the baseband signal received through the channel characterized by $h^\star\mathrm{(}\tau, \upsilon\mathrm{)}$, with both $s(t)$ and $r^\star(t)$ being functions of the time variable $t$.
	
	\subsection{Signal Model}
	Considering the significant DD spreading characteristic of the LEO satellite channels, we apply the ODDM waveform to the integrated communication and remote sensing system.
	
	\subsubsection{Transmitted ODDM Signal Model}
	The ODDM modulator places $M\times N$ input complex digital symbols in the DD plane with certain delay resolution $\mathcal{T}$ and Doppler resolution $\mathcal{F}$ as a matrix $\mathbf{X}\left[m,n\right]$, $m\in\{0,1,\dots, M-1\}$ and $n\in\{0,1,\dots, N-1\}$. These DD-domain symbols are converted to time-domain signals by applying multi-carrier modulation with a DDOP $g\mathrm{(}t\mathrm{)}$ as
	\begin{equation} \label{eq9}
		s(t)=\sum^{M-\mathrm{1}}_{m\mathrm{=0}}{}\sum^{\frac{N}{2}-1}_{n=-\frac{N}{2}}\mathbf{X}\left[m,(n)_N\right]g(t-m\mathcal{T})e^{j\mathrm{2}\pi n\mathcal{F}(t-m\mathcal{T})},
	\end{equation}
	where $g\mathrm{(}t\mathrm{)}$ is a pulse training sequence, which maintains orthogonality with respect to the DD plane’s resolutions, and has a joint DD resolution $\mathcal{R}=\mathcal{T}\mathcal{F}=\frac{1}{MN}$. Specifically, $g\mathrm{(}t\mathrm{)}$ can be represented as
	\begin{equation} \label{eq10}
		g\mathrm{(}t\mathrm{)=}\sum^{N-\mathrm{1}}_{\dot{n}\mathrm{=0}}{}a\mathrm{(}t-\dot{n}T_0\mathrm{)},
	\end{equation}
	where $a\mathrm{(}t\mathrm{)}$ is a square-root Nyquist subpulse with a zero-ISI interval of $\frac{T_0}{M}$ and a pulse duration of $\mathrm{2}D\frac{T_0}{M}$, where $D$ is a positive integer. $T_0$ is the subpulse spacing of $g(t)$, given as $T_0=M\mathcal{T}=\frac{\mathrm{1}}{N\mathcal{F}}$.
	
	In general, the ODDM modulation procedure described above can be approximately implemented in a matrix form as \cite{cite24}
	\begin{equation} \label{eq11}
		\mathbf{s} = \mathrm{vec}(\mathbf{X}\mathbf{F}^{\mathrm{H}}_N),
	\end{equation}
	where $\mathbf{s} \in \mathbb{C}^{MN \times 1}$ is a time-domain sequence with a sampling interval $T_s = \frac{T_0}{M}$. It is worth pointing out that, when $M = 1$, the symbol matrix $\mathbf{X}\left[m,n\right]$ degenerates into a frequency-domain symbol vector. In this case, the ODDM modulation procedure described by \eqref{eq11} can be regarded as OFDM modulation.
	
	Then, a CP of length $L$ is appended to $\boldsymbol{\mathrm{s}}$, where $L$ corresponds to the maximum number of path delay difference, i.e., $L = \left\lceil \frac{{\tau_{\max} - \tau_{\min}}}{T_s} \right\rceil$, where $\lceil\cdot\rceil$ denotes the ceiling operator.
	Afterward, the sequence $\boldsymbol{\mathrm{s}}$ with the added CP is passed through pulse shaping with $a(t)$ in \eqref{eq10} to obtain the baseband ODDM signal $s(t)$.
	The baseband signal is upconverted to radio frequency (RF) and transmitted through the communication channel $h^\mathrm{c}(\tau,\upsilon)$ to user devices, and simultaneously through the round-trip SAR channel $h^\mathrm{r}(\tau,\upsilon)$ back to the satellite receiver.
	
	\subsubsection{Received ODDM Signal Model}
	Generally, the ODDM receiver downconverts the received RF signals to obtain baseband signal $r^\star(t)$ as described in \eqref{eq8}. Next, $r^\star(t)$ undergoes matched filtering with $a(t)$, performed by an analog-to-digital converter (ADC), to obtain the received time-domain sequence $\boldsymbol{\mathrm{r}}^\star\in\mathbb{C}^{MN\times 1}$, where the symbol rate $\mathrm{W}=\frac{M}{T_0}$ and total duration $\mathrm{T}=NT_0$.
	The receiver then applies column-wise devectorization and row-wise $N$-point discrete Fourier transform to demodulate $\boldsymbol{\mathrm{r}}^\star$ into DD-domain samples as a matrix $\mathbf{Y}^\star\left[m,n\right]$ for $m\in\{0,1,\dots, M-1\}$ and $n\in\{0,1,\dots, N-1\}$, which can be represented by
	\begin{equation} \label{eq12}
		\mathbf{Y}^\star=\mathrm{vec}^{-1}_{M,N}(\mathbf{r}^\star){\mathbf{F}}_{N}+{\mathbf{Z}},
	\end{equation}
	where $\mathrm{vec}^{-1}_{M,N}(\cdot)$ denotes column-wise devectorization operator, which transforms an $MN$-length vector back into an $M\times N$ matrix, and $\boldsymbol{\mathrm{Z}}\in \mathbb{C}^{M\times N}$ is the additive white Gaussian noise (AWGN) matrix composed of DD-domain noise samples with variance ${\sigma }^{\mathrm{2}}$.
	
	Since $\boldsymbol{\mathrm{Y}}^\star$ is significantly distorted by the DD spreading of the channel, pilot symbols need to be embedded among the input DD-domain symbols for effective channel sensing. Consequently, the detectable channel parameters are confined to the same DD plane with delay resolution $\mathcal{T}=\frac{T_0}{M}$ and Doppler resolution $\mathcal{F}=\frac{1}{NT_0}$. Let the maximum delay difference and Doppler shift of the channel be $L\frac{T_0}{M}$ and $K\frac{1}{NT_0}$, respectively. It is reasonably assumed that the satellite's beam has a forward-looking squint, thus resulting in $K$ being a positive integer. To represent the full set of DD-domain channel parameters, we define a general DD channel matrix $\boldsymbol{\mathrm{H}}^\star \in \mathbb{C}^{\left(L+1\right) \times \left(K+1
	\right)}$ in the DD plane, where each row and each column of $\boldsymbol{\mathrm{H}}^\star$ correspond to a delay and Doppler tap, respectively, i.e., delay tap $l \in \{0, 1, \dots, L\}$ and Doppler tap $k \in \{0, 1, \dots, K\}$.
	With the general channel matrix $\boldsymbol{\mathrm{H}}^\star$, the exact DD-domain IO relation between $\mathbf{X}$ and $\mathbf{Y}^\star$ can be effectively represented by
	\begin{align} \label{eq13}
		\mathbf{Y}^\star\left[m,n\right]
		&= \sum^{L}_{l=0} \sum^{K}_{k=0} \mathbf{H}^\star[l, k] \mathbf{X}\left[(m-l)_M, (n-k)_N\right] \nonumber \\
		&\quad \cdot e^{j 2\pi \frac{k(m-l)}{MN}} {\gamma}_{m,n} + \mathbf{Z}[m,n],
	\end{align}
	where ${\gamma }_{m,n}$ is a phase factor given as
	\[{\gamma }_{m,n}\mathrm{=}\left\{ \begin{array}{ll}
		\mathrm{1} & l\le m\le M-1 \\
		e^{-j\mathrm{2}\pi \frac{\mathrm{[}n-k{\mathrm{]}}_N}{N}} & 0\le m\mathrm{<}l \end{array}
	\right.. \]
	Depending on the specific channel type,  $\mathbf{H}^\star$ can be formulated as follows.
	
	When $\star=\mathrm{c}$, $\mathbf{H}^\mathrm{c}$ involves channel parameters associated with the communication channel in \eqref{eq4}. The delay ${\tau }_p$ and Doppler shift $\upsilon_p$ of the $p$-th path of the communication channel are on the DD grid, i.e., ${\tau }_p\mathrm{=}l_p\frac{T_0}{M}$ and ${\upsilon }_p\mathrm{=}\frac{k_c+\kappa_p}{NT_0}$,
	where $l_p$	represents integer delay tap of the $p$-th path, $k_c$ denotes a common integer Doppler tap induced by the mobility of the LEO satellite, and $\kappa_p$ correspond to the fractional component of the Doppler tap.
	The differing Doppler shifts of multiple paths lead to significant time selectivity in the channel. However, apart from the common Doppler shift $\upsilon_{\scriptscriptstyle \mathrm{Sat}}$ resulted by the LEO satellite, this effect is typically negligible when the signal duration remains within the channel coherence time. Consequently, the Doppler shift for each path can be approximated as consistent, with the fractional part of the Doppler shift compensated using a CP-based carrier frequency offset (CFO) estimator \cite{cite27}. Hence, the communication channel matrix $\mathbf{H}^\mathrm{c}\left[l,k\right]$ can be effectively represented by
	\begin{equation} \label{eq14}
		\mathbf{H}^\mathrm{c}[l,k] =
		\begin{cases}
			\alpha_p e^{-j2\pi f_0 \tau_p}, & \text{if } (l = l_p) \wedge (k = k_c) \\
			0, & \text{otherwise}
		\end{cases}.
	\end{equation}

	When $\star=r$, $\mathbf{H}^\mathrm{r}$ refers to channel parameters for the SAR remote sensing channel in \eqref{eq6}. To synchronize with the beginning of the received signal, the ADC initiates sampling after $\frac{2R_0(\eta)}{c}$. Then $\boldsymbol{\mathrm{H}}^\mathrm{r}\left[l,k\right]$ can be similarly represented by
	\begin{equation} \label{eq15}
		\mathbf{H}^\mathrm{r}[l,k] =
		\begin{cases}
			\beta_q e^{-j2\pi f_0 \tilde{\tau}_q}, & \text{if } (l = q) \wedge (k = k_r) \\
			0, & \text{otherwise}
		\end{cases}.
	\end{equation}
	where $k_r$ is the integer Doppler tap corresponding to the intra-pulse Doppler shift $\tilde{\upsilon}_r$, with the fractional part compensated through CP-based processing.
	
	\subsection{Work Mode}
\begin{figure}[tbph!]
		\centering
		\includegraphics[width=1\linewidth]{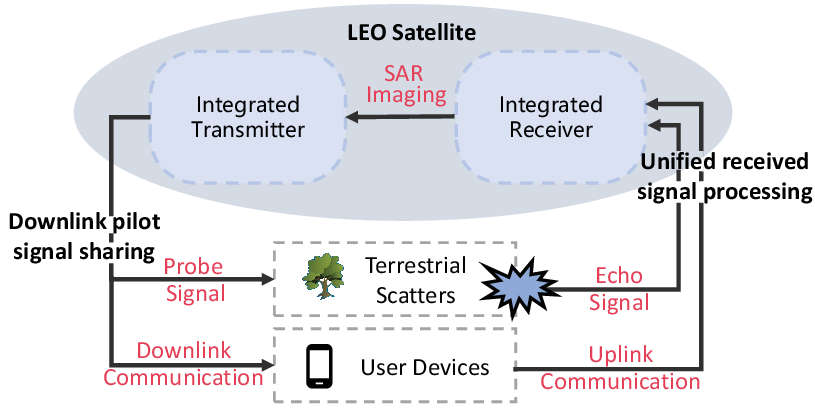}
		\captionsetup{labelformat=empty}
		\caption{\small Fig.~3.~~Work mode for integrated communication and remote sensing system.}
		\label{fig3}
	\end{figure}

	Generally, the integrated communication and remote sensing system needs to support two-way communications between LEO satellite and user devices. For example, the user device sends remote sensing request through uplink communication, while the LEO satellite transmits remote sensing data via downlink communications. In the following, we introduce the work mode as described in Fig. \ref{fig3}.
	
	The integrated communication and remote sensing system works in a time duplex division (TDD) mode. In other words, a wireless frame divides into downlink and uplink communication time slots. During downlink communication times slots, the LEO satellite transmits signals and receives the echo signals reflected by terrestrial scatters, while the user devices also receive and decode the downlink signals. During uplink communication time slots, the terrestrial user devices send signals to the LEO satellite. The allocation of time slots for downlink and uplink can be conducted according to the requirements of considered scenarios.
	
	To realize the described work mode, a specific transmission protocol and transceiver processing architecture are proposed to support integrated communication and remote sensing, as detailed in Sections III and IV, respectively. The integrated framework is further implemented on an SDR platform in Section V to validate its real-time SAR imaging and information transmission capabilities.
	\section{Protocol Design for Integrated Communication and Remote Sensing}
	\label{sec3}
	In this section, we propose a transmission protocol to effectively support integrated communication and SAR remote sensing. Specifically, we first provide a flexibly adjustable frame structure to enable the sharing of downlink pilot signals under PRF restrictions in different LEO satellite configurations. Next, we design a symbol pattern for the pilot signal to facilitate unified channel sensing.
	\begin{figure*}[tbph!]
		\centering
		\includegraphics[width=0.88\linewidth]{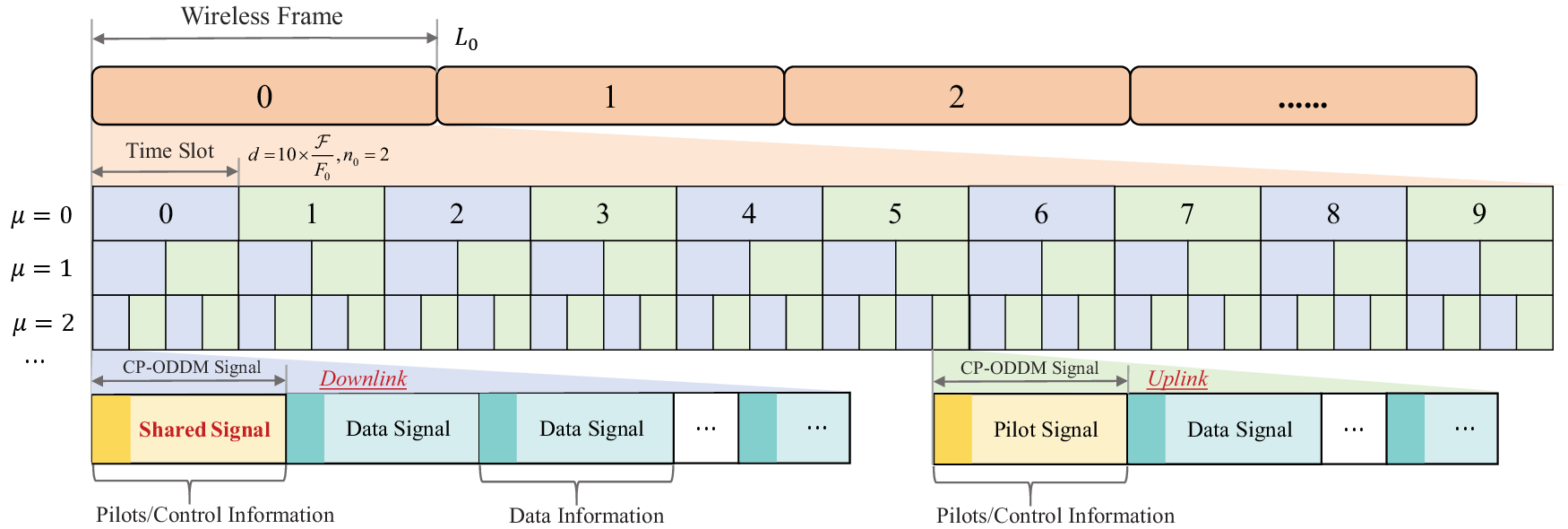}
		\captionsetup{labelformat=empty}
		\caption{\small Fig.~4.~~Integrated wireless frame structure based on 5G NR standard.}
		\label{fig4}
	\end{figure*}
	
	\subsection{Wireless Frame Structure}
	According to the characteristics and requirements of integrated communication and SAR remote sensing in LEO satellite systems, we design a wireless frame structure as depicted in Fig. \ref{fig4}. Notice that such a frame is designed based on the upper layer frame structure of 5G NR standard for compatibility, and applies ODDM as the physical layer waveform to efficiently adapt to the high-mobility LEO satellite environment. Specifically, the length of each wireless frame is fixed at $L_0$. Each frame comprises $d=10\times 2^{\mu }$ time slots of equal length, where $\mu$ is a numerology index relevant to Doppler resolution $\mathcal{F}$ of ODDM modulation, i.e., $\mathcal{F}=2^{\mu }F_0$, with $F_0$ being a reference frequency constant. In other words, each time slot has a length equal to an $N_s$ CP-ODDM signals where the upper bound is given by $\left\lfloor\frac{L_0}{10}F_0\right\rfloor$, with $\lfloor\cdot\rfloor$ denoting the flooring operator. Finally, the number of modulation symbols constituting an ODDM signal is proportional to its bandwidth, i.e., $B_w = MN \cdot \mathcal{F}$.
	
	In the designed wireless frame, shared signals denote downlink pilot signals adopted as both pilot sequences in communication and probe signals for SAR imaging. Notably, the probe signals of SAR need to be transmitted at a restricted PRF to prevent range and azimuth ambiguity, i.e., $\mathrm{PR}{\mathrm{F}}_{\min}\mathrm{=}\frac{V_s}{{\rho }_a}\cos\mathrm{(}{\theta }_s\mathrm{)}$ and $\mathrm{PR}{\mathrm{F}}_{\max}\mathrm{=(}\frac{\mathrm{2}R_w}{c}\cos\mathrm{(}{\theta }_g\mathrm{)}+\frac{L_0}{dN_s}{\mathrm{)}}^{-\mathrm{1}}$, where ${\rho }_a\mathrm{=}\frac{\mathrm{0.886c}R_c}{{2f}_0R_a}$ denotes the azimuth resolution and $R_a=V_sT_s$ is the synthetic aperture length. Therefore, the shared signals need to be periodically embedded in the frame structure according to the restricted PRF of specific LEO satellite configuration.
	To facilitate the PRF management, let the minimum period for the downlink slot to appear in the frame structure be denoted as $n_0$, with pilot signals assigned at the beginning of every $n_0$-time slot (in the example of Fig. \ref{fig4}, the first ODDM signal of every $n_0\mathrm{=2}$ time slots serves as shared downlink pilot signals). Then, the proposed frame structure, which inherits the flexibility of the 5G NR frame structure, allows the PRF to be controlled by adjusting the numerology index $\mu$ according to the specific LEO satellite configuration, i.e., $\mathrm{PRF}\mathrm{=}\frac{\mathrm{10}\mathrm{\times }2^{\mu }}{n_0L_0}$, and the lower boundary value of the numerology index $\mu$ can be directly determined by satellite system parameters as
	\begin{equation} \label{eq16}
		\mathrm{\mu }\geq\left\lceil {\log}_{\mathrm{2}}\left(\frac{{n_0L}_0{V}_{s}}{\mathrm{10}{\mathrm{\rho}}_{a}}\mathrm{cos(}{\theta }_{s}\mathrm{)}\right)\right\rceil.
	\end{equation}
	
	Thus, the proposed wireless frame structure is flexibly adjustable to ensure the efficient PRF management for SAR imaging under different LEO satellite configurations.
	
	\subsection{Symbol Pattern for Pilot Signals}
	As mentioned earlier, the ODDM demodulator mitigates fractional Doppler effects by controlling the length of the ODDM signal within the coherent time and employing a CP-based CFO estimator. However, the impact of fractional delay still exists, causing delay tap spreading in the DD plane. To address this problem, a line of pilots is embedded in DD plane for DD-domain channel sensing. To improve the efficiency, the pilots and control information are arranged together in the DD plane as a pilot ODDM signal.
	Moreover, the downlink pilot signal for sharing is applied with a sufficiently long CP to address the large delay differences caused by the SAR channel, and transmitted with higher power to mitigate the power loss of round-trip propagation.
	\begin{figure}[tbph!]
		\centering
		\includegraphics[width=1\linewidth]{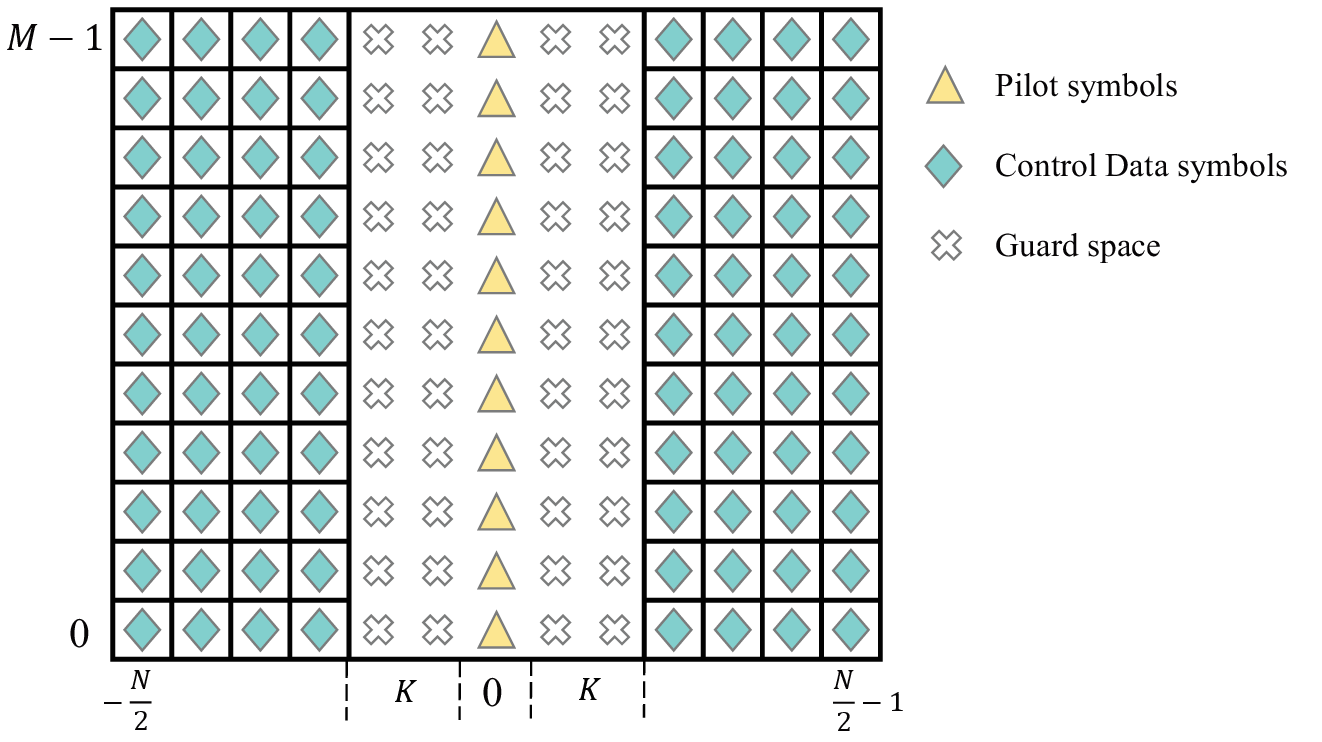}
		\captionsetup{labelformat=empty}
		\caption{\small Fig.~5.~~Symbol pattern for pilot signals.}
		\label{fig5}
	\end{figure}
	
	Let the symbols of pilots and control information be arranged together in matrix ${\boldsymbol{\mathrm{X}}}^{\mathrm{S}}\left[m,n\right]\in \mathbb{C}^{M\times N}$ to form the pilot signal. As shown in Fig. \ref{fig5}, a vector of pilot symbols $\boldsymbol{\mathrm{u}}\mathrm{=}\mathrm{\{}u_m{\}}^{M-\mathrm{1}}_{m\mathrm{=0}}$ is inserted in the ${\boldsymbol{\mathrm{X}}}^{\mathrm{S}}$. Additionally, the guard space used for mitigating interference from control information symbols to pilot symbols is arranged in the range $m\in \{\mathcal{M}_g|0\leq m \leq M-1\}$ and $n\in \{\mathcal{N}_g|-K\leq n \leq K, n\neq0\}$.
	Then the whole DD-domain symbols can be expressed with the pilot symbols $\mathbf{u}$ and the control information symbols $\mathbf{X}\left[m, n\right]$ as
	\begin{equation} \label{eq17}
		{\boldsymbol{\mathrm{X}}}^{\mathrm{S}}\left[m,n\right]\mathrm{=}\left\{ \begin{array}{ll}
			u_m, & n\mathrm{=0} \\
			0, & n\in \mathcal{N}_g \\
			{\boldsymbol{\mathrm{X}}}\left[m,n\right], & \mathrm{otherwise}
			\end{array}
		\right..
	\end{equation}
	
	In 5G NR, Zadoff-Chu (ZC) sequence is adopted as the pilot symbol vector $\boldsymbol{\mathrm{u}}$ \cite{cite28}. Assuming that control information symbols are absent in the pilot signal, the constant modulus characteristic of the ZC sequence is preserved in the time domain under ODDM modulation procedure in \eqref{eq11}. This preservation ensures that the shared signal maintains a low peak-to-average power ratio (PAPR). Moreover, since the symbol pattern of the pilots is intricately related to the specific channel sensing scheme employed, we will further discuss and analyze the corresponding channel sensing algorithm in the next section. It will be confirmed that the minimum mean square error (MSE) performance of channel sensing is achieved when ZC sequence is utilized as the pilot symbols. Hence, accurate remote sense can also be achieved.
	
	\section{Architecture Design for Integrated Communication and Remote Sensing}
	\label{sec4}
	In this section, we design a system architecture to effectively support integrated communication and SAR remote sensing. Moreover, we propose a low-complexity but high-performance channel sensing algorithm for realizing integrated communication and remote sensing.
	\begin{figure*}[tbph!]
		\centering
		\includegraphics[width=0.88\linewidth]{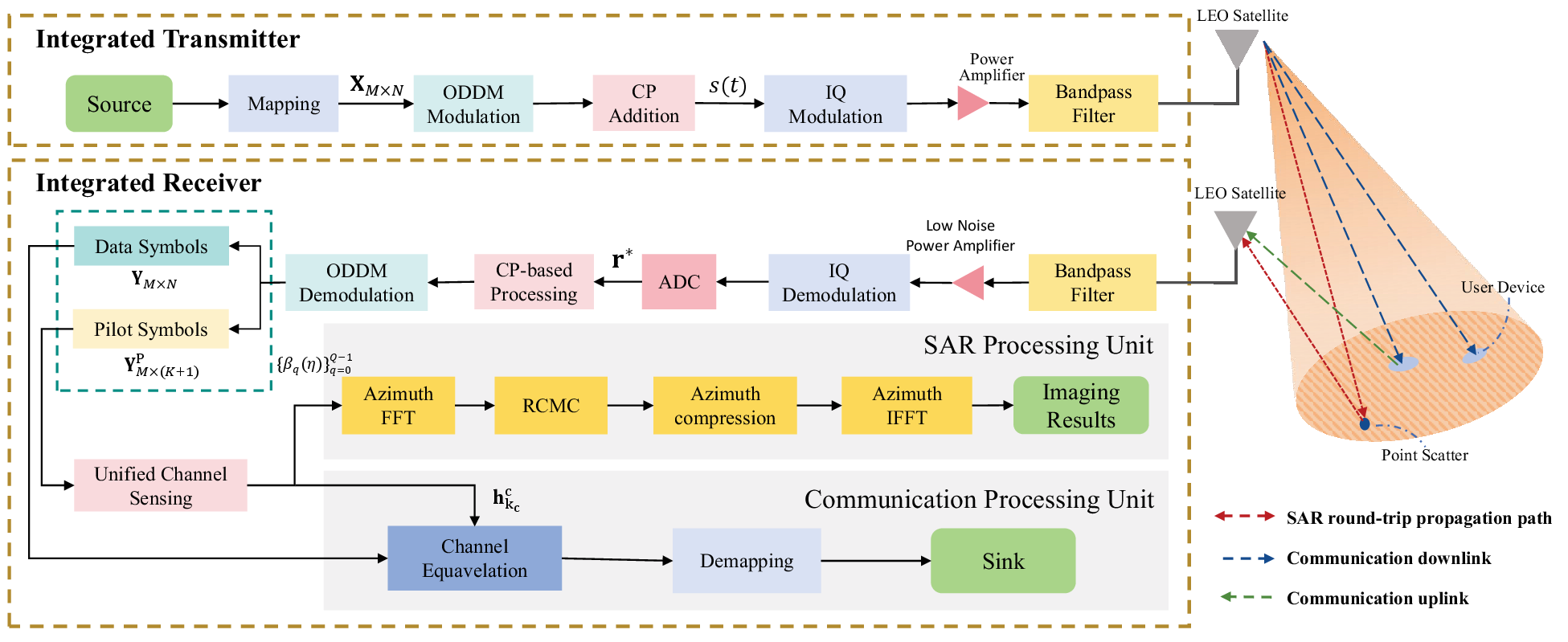}
		\captionsetup{labelformat=empty}
		\caption{\small Fig.~6.~~System architecture for integrated communication and SAR remote sensing.}
		\label{fig6}
	\end{figure*}
	
	As depicted in Fig. \ref{fig6}, the system architecture includes an integrated transmitter and an integrated receiver. At the integrated transmitter, signals undergo mapping, ODDM modulation, CP addition, in-phase and quadrature (IQ) modulation, power amplification, bandpass filter, and then are transmitted. Data signals facilitate flexible resource allocation through controllable DD-domain source symbols, while pilot signals, with a designed symbol pattern, utilize pilot symbols for channel sensing and other symbols as control information.
	
	At the integrated receiver, the received signals undergo bandpass filter, low noise amplification, IQ demodulation, ADC, CP-based processing and ODDM demodulation, then become DD-domain data symbols and pilot symbols. The dispersed pilot symbols are processed for channel sensing, enabling both channel estimation for communication and range reconstruction for SAR imaging. With the estimated channel state information (CSI), minimum mean square error (MMSE) channel equalization is performed over data symbols, while SAR imaging is achieved using conventional SAR processing methods, including range cell migration correction (RCMC) and azimuth compression. In the following, a unified signal processing framework is designed for the integrated receiver, including channel sensing, channel equalization for communication and range reconstruction for SAR imaging.

	\subsection{Unified Channel Sensing}
	Herein, we propose a unified channel sensing algorithm that utilizes the received pilot symbols to extract the DD channel matrix $\mathbf{H}^\star$. For clarity, we define $h^\star_{l,k}=\mathbf{H}^\star\left[l,k\right]$, for $l\in\{0,1,\dots,L\}$ and $k\in\{0,1,\dots,K\}$.
	
	Due to the DD spreading, the received pilot symbols ${\boldsymbol{\mathrm{Y}}}^{\mathrm{P}}\left[m,n\right]\in \mathbb{C}^{M\times (K+1)}$ may be diffused in the DD plane, which can be represented by substituting \eqref{eq17} into \eqref{eq13} as
	\begin{equation} \label{eq18}
		{\boldsymbol{\mathrm{Y}}}^{\mathrm{P}}\left[m,n\right]\mathrm{=}\sum^{L}_{l=0}h^\star_{l,n}u_{\mathrm{[}m-l{\mathrm{]}}_M}e^{j\mathrm{2}\pi \frac{\tilde{n}\mathrm{(}m-l\mathrm{)}}{MN}}+{\mathbf{Z}}\left[m,n\right] ,
	\end{equation}
	where $\tilde{n}\mathrm{=[}n+\frac{N}{\mathrm{2}}{\mathrm{]}}_N-\frac{N}{\mathrm{2}}$, and the $h^\star_{l,k}$ has Doppler tap $k=n$.
	
	Depart the received pilot symbols ${\boldsymbol{\mathrm{Y}}}^{\mathrm{P}}$ into $M\times \mathrm{1}$ vectors ${\boldsymbol{\mathrm{y}}}^{\mathrm{P}}_{n}$, $n\in \{0,\dots, K\}$, i.e., ${\boldsymbol{\mathrm{Y}}}^{\mathrm{P}}=\left[{\boldsymbol{\mathrm{y}}}^{\mathrm{P}}_{\mathrm{0}}, {\boldsymbol{\mathrm{y}}}^{\mathrm{P}}_{\mathrm{1}}, \dots, {\boldsymbol{\mathrm{y}}}^{\mathrm{P}}_{K}\right]$. Then the relation between the received vectors and transmitted pilot vector can be expressed as
	\begin{equation} \label{eq19}
		{\mathbf{y}}^{\mathrm{P}}_{n}={\mathbf{H}}^\mathrm{P}_{n}{\mathbf{C}}_{n}{\mathbf{u}}+{\mathbf{z}}_{n},
	\end{equation}
	where ${\mathbf{z}}_{n}$ is the $n$-th column of DD-domain noise matrix $\mathbf{Z}$,  ${\mathbf{C}}_{n}\mathrm{=diag(1,}W^{-\tilde{n}}_{MN},W^{-\mathrm{2}\tilde{n}}_{MN},\dots ,W^{-\mathrm{(}M-\mathrm{1)}\tilde{n}}_{MN}\mathrm{)}$, and the $M\times M$ channel matrix ${\mathbf{H}}^\mathrm{P}_{n}$ is given by
	\begin{equation} \label{eq20}
		{\boldsymbol{\mathrm{H}}}^\mathrm{P}_{n}\mathrm{=}\left( \begin{array}{llll}
			h^\star_{0,n} &
			\cdots  &
			h^\star_{2,n}W^{n}_N &
			h^\star_{1,n}W^{n}_N \\
			h^\star_{1,n} &
			h^\star_{0,n} &
			\ddots  &
			h^\star_{\mathrm{2,}n}W^{n}_N \\
			\vdots  & \ddots  & \ddots & \vdots  \\
			h^\star_{M-1,n} & \cdots & h^\star_{1,n} & h^\star_{0,n} \end{array}
		\right).
	\end{equation}
	Since ${\boldsymbol{\mathrm{H}}}^\mathrm{P}_{n}$ is factor-circulant, it can be represented as \cite{cite23}
	\begin{equation} \label{eq21}
		{\mathbf{H}}^\mathrm{P}_{n}={\mathbf{C}}_{n}{\mathbf{F}}^{\mathrm{H}}_{M}{\mathbf{D}}^\mathrm{P}_{n}{\mathbf{F}}_{M}{\mathbf{C}}^{-1}_{n},
	\end{equation}
	where ${\mathbf{D}}^\mathrm{P}_{n}=\mathrm{diag}(\sqrt{M}{\mathbf{F}}_{M}{\mathbf{C}}^{-1}_{n}{\mathbf{h}}^\star_{n})$, in which ${\mathbf{h}}^\star_{n}\in \mathbb{C}^{M \times 1}$ is the zero-padded augmented vector of the $n$-th column of the DD channel matrix $\mathbf{H}^\star$, i.e., ${\mathbf{h}}^\star_n = \left[ \mathbf{H}^\star[0, n], \dots, \mathbf{H}^\star[K, n], \mathbf{0}  \right]^\mathrm{T}$. By substituting \eqref{eq21} into \eqref{eq19}, we have
	\begin{equation} \label{eq22}
		\begin{array}{ll}
			{\widehat{\mathbf{h}}}^\star_{n} & =\frac{1}{\sqrt{M}}{\mathbf{C}}_{n}{\mathbf{F}}^{\mathrm{H}}_{M}\left({\mathbf{F}}_{M}{\mathbf{C}}^{-1}_{n}{\mathbf{y}}^{\mathrm{P}}_{n}\oslash{\mathbf{F}}_{M}\mathbf{u}\right) \\
			& ={\mathbf{h}}^\star_{n}+{\mathbf{z}}_{n}' \end{array},
	\end{equation}
	where $\oslash$ denotes the Hadamard division operation, and  ${\mathbf{z}}_{n}'$ is an $M\times 1$ noise vector with improved variance $\frac{{\sigma }^{2}}{M^{2}}\sum^{M-1}_{m=0}{\left|{\tilde{u}}_m\right|}^{-2}$, where ${\tilde{u}}_m$ is the $m$-th element of ${\mathbf{F}}_{M}\mathbf{u}$. Thus, we obtain the $M\times (K+1)$ estimated CSI matrix $\widehat{\mathbf{H}}^\star=\left[{\widehat{\mathbf{h}}}^\star_0,{\widehat{\mathbf{h}}}^\star_{1},\dots ,{\widehat{\mathbf{h}}}^\star_{K}\right]$.
	
	\subsection{Channel Equalization for Communication}
	Then, we propose a low-complexity MMSE equalization method to recover data symbols from distorted DD-domain samples with the estimated CSI matrix ${\widehat{\mathbf{H}}}^\mathrm{c}$.
	
	We first reformulate \eqref{eq13} in a matrix form with transmitted data symbol vectors $\mathbf{x}_n=\big[\mathbf{X}[0,n], \mathbf{X}[1,n], \dots, \mathbf{X}[M-1,n]\big]^\mathrm{T}$ and received DD-domain sample vectors $\mathbf{y}_n=\big[\mathbf{Y}[0,n], \mathbf{Y}[1,n], \dots, \mathbf{Y}[M-1,n]\big]^\mathrm{T}, n\in\{0,\dots,N-1\}$ as
	\begin{equation}\label{eq23}
		\mathbf{y}_n=\mathbf{C}_n\mathbf{F}_M^\mathrm{H}\mathbf{D}_n\mathbf{F}_M\mathbf{C}_n^\mathrm{-1}\mathbf{C}_{k_c}\mathbf{x}_{(n-k_c)_N}+\mathbf{z}_{n},
	\end{equation}
	where $\mathbf{D}_n=\mathrm{diag}(\sqrt{M}\mathbf{F}_M\mathbf{C}^{-1}_{n}\mathbf{h}^{\mathrm{c}}_{k_c})$. Define $\widetilde{\mathbf{y}}_n=\mathbf{F}_M\mathbf{C}^\mathrm{-1}_n\mathbf{y}_n$ and $\widetilde{\mathbf{H}}_{n}=\mathbf{D}_{n}\mathbf{F}_{M}\mathbf{C}_{n}^\mathrm{-1}\mathbf{C}_{k_c}$, then \eqref{eq23} can be simplified as
	\begin{equation} \label{eq24}
		\widetilde{\mathbf{y}}_{n}=\widetilde{\mathbf{H}}_{n}\mathbf{x}_{(n-k_c)_N}+\mathbf{z}_{n}.
	\end{equation}
	
	Then, apply an MMSE estimation matrix $\mathbf{W}\in\mathbb{C}^{M\times M}$ to recover transmitted symbols, i.e., $\widehat{\mathbf{x}}_{(n-k_c)_N}=\mathbf{W}\widetilde{\mathbf{y}}_{n}$. Therefore, the MMSE problem can be established as
	\begin{equation} \label{eq25}
		\mathbf{W}=\arg\min\limits_{\substack{\mathbf{W}}} \mathbb{E}\{{\big\lVert \mathbf{W}\widetilde{\mathbf{y}}_{n}-\mathbf{x}_{(n-k_c)_N}\big\rVert}^2\}.
	\end{equation}
	
	By solving problem \eqref{eq25}, the MMSE estimation matrix $\mathbf{W}$ can be calculated as
	\begin{align} \label{eq26}
		\mathbf{W} &= \widetilde{\mathbf{H}}_{n}^\mathrm{H} \left( \widetilde{\mathbf{H}}_{n} \widetilde{\mathbf{H}}_{n}^\mathrm{H} + \sigma^2 \mathbf{I}_M \right)^{-1} \notag \\
		&= \mathbf{C}_{k_c}^{-1} \mathbf{C}_{n} \mathbf{F}_{M}^\mathrm{H} \mathbf{D}_{n}^\mathrm{H} {\left( \mathbf{D}_{n} \mathbf{D}_{n}^\mathrm{H} + \sigma^2 \mathbf{I}_M \right)}^{-1}.
	\end{align}
	
	Since $\mathbf{D}_{n}$ can be estimated with $\widehat{\mathbf{h}}_{k_c}$ as a diagonal matrix, the inverse of $(\mathbf{D}_{n} \mathbf{D}_{n}^\mathrm{H} + \sigma^2 \mathbf{I}_M)$ can be computed by element-wise division of $\widetilde{\mathbf{y}}_{n}$ with $\mathrm{diag}^{-1}(\mathbf{D}_{n} \mathbf{D}_{n}^\mathrm{H} + \sigma^2 \mathbf{I}_M)$ as
	\begin{equation} \label{eq27}
		\acute{\mathbf{y}}_{n}=\widetilde{\mathbf{y}}_{n}\oslash \mathrm{diag}^{-1}(\mathbf{D}_{n} \mathbf{D}_{n}^\mathrm{H} + \sigma^2 \mathbf{I}_M),
	\end{equation}
	where $\mathrm{diag}^{-1}(\cdot)$ denotes the operator that transfers diagonal matrix into a vector. Then, the recovered symbol vector $\widehat{\mathbf{x}}_{(n-k_c)_N}$ can be effectively expressed as
	\begin{equation} \label{eq28}
		\widehat{\mathbf{x}}_{(n-k_c)_N}=\mathbf{C}_{k_c}^{-1} \mathbf{C}_{n} \mathbf{F}_{M}^\mathrm{H} \mathbf{D}_{n}^\mathrm{H}\acute{\mathbf{y}}_{n}.
	\end{equation}
	
	Notably, since the multiplication by $\mathbf{F}_{M}$ can be efficiently performed using the fast Fourier transform (FFT), the proposed MMSE equalization method exhibits low computational complexity, making it well-suited for LEO satellite systems.
	
	\subsection{Range Reconstruction for SAR Imaging}
	Under the proposed wireless frame structure, the shared downlink pilot signals can be regarded as periodically transmitted probe signals. Therefore, the echo signals through the SAR channel can be expressed in terms of range time $t$ and azimuth time $\eta$ as
	\begin{equation} \label{eq29}
		r(t,\eta)=\sum^{Q-\mathrm{1}}_{q\mathrm{=0}}{}\beta_q\mathrm{(}\eta \mathrm{)}e^{-j\mathrm{2}\pi f_0\tilde{\tau}_q}s\left(t-\tilde{\tau}_q\right).
	\end{equation}
	
	With the ADC in the satellite receiver initiating sampling after $\frac{2R_0(\eta)}{c}$, the reconstructed range profile can be derived from the continuous set of $Q$ values contained in the estimated CSI vector  ${\widehat{\mathbf{h}}}^{\mathrm{r}}_{k_r}$ at each azimuth time $\eta$, denoted as ${\{\widehat{\beta}_q(\eta)\}}_{q=0}^{Q-1}$. After the equivalent range reconstruction, this two-dimensional range-compressed signal can be expressed as
	\begin{equation} \label{eq30}
		{r}_{rc}(t, \eta) = \sum_{q=0}^{Q-1} \widehat{\beta}_q(\eta)\delta\left(t - \frac{2 R_q(\eta)}{c}\right),
	\end{equation}
	which is in a standard format, allowing the application of conventional SAR processing methods, including RCMC and azimuth compression \cite{cite26}, to obtain the final imaging results. Nevertheless, in spaceborne SAR scenarios, phase errors arising from non-ideal satellite trajectories inevitably lead to image defocusing and severely impair the imaging quality. To address this, post-processing techniques such as phase gradient autofocus (PGA) \cite{303752} and minimum-entropy autofocus (MEA) \cite{wang2006sar} are essential after the SAR image formation stage to ensure accurate focusing.
	
	Collectively, the implementation steps for each component of the proposed integrated receiver are cohesively summarized in \textbf{Algorithm 1}.
	\begin{algorithm}
		\caption{Integrated communication and remote sensing algorithm}
		\begin{algorithmic}[1] 
			\STATE \textbf{Precomputed:} $(\mathbf{F}_M)_{M\times M}$, $\left\{ (\mathbf{C}_{n})_{M\times M} \right\}_{n=0}^{N-1}$
			\STATE \textbf{Unified channel sensing:}
			\STATE \hspace{1em} \textbf{Input:} transmitted pilot symbols $\mathbf{u}_{M\times 1}$, \\
			\hspace{1em} received DD-domain pilot samples $\mathbf{Y}^\mathrm{P}_{M\times (K+1)}$
			\STATE \hspace{1em} \textbf{Output:} DD channel state information $\widehat{\mathbf{H}}^\star_{M\times (K+1)}$.
			\STATE \hspace{1em} \textbf{for}~$n=0$~\text{to}~$K$~\textbf{do}
			\STATE \hspace{2em} Compute: $(\mathbf{y}_{n}^\mathrm{P})_{M\times 1} \leftarrow n$\text{-th column of} $\mathbf{Y}^\mathrm{P}$
			\STATE \hspace{2em} Compute: $\widehat{\mathbf{D}}^\mathrm{P}_{n}\leftarrow(\mathbf{F}_{M}\mathbf{C}_{n}\mathbf{y}^\mathrm{P}_{n})_{M\times1}\oslash (\mathbf{F}_M\mathbf{u})_{M\times1}$
			\STATE \hspace{2em} Compute: $(\widehat{\mathbf{h}}_{n}^\star)_{M\times 1}\leftarrow \mathbf{C}_n\mathbf{F}_{M}^\mathrm{H}\widehat{\mathbf{D}}^\mathrm{P}_{n}$
			\STATE \hspace{1em} \textbf{end for}
			\STATE \hspace{1em} \textbf{return}~$\widehat{\mathbf{H}}^\star_{M\times (K+1)}=[\mathbf{\widehat{h}}^\star_0, \mathbf{\widehat{h}}^\star_1, \dots, \mathbf{\widehat{h}}^\star_{K}]$
			
			\vspace{0.2em} 
			\STATE \textbf{Channel equalization:}
			\STATE \hspace{1em} \textbf{Input:} estimated communication channel gains $\widehat{\mathbf{H}}^\mathrm{c}$, \\
			\hspace{1em} received DD-domain data samples $\mathbf{Y}_{M\times N}$
			\STATE \hspace{1em} \textbf{Output:} recovered DD-domain data symbols $\widehat{\mathbf{X}}_{M\times N}$.
			\STATE \hspace{1em} \textbf{for}~$n=0$~\text{to}~$N-1$~\textbf{do}
			\STATE \hspace{2em} Compute: $(\mathbf{y}_{n})_{M\times 1} \leftarrow n$\text{-th column of} $\mathbf{Y}$
			\STATE \hspace{2em} Compute: $(\mathbf{D}_{n})_{M\times M}\leftarrow\mathrm{diag}(\mathbf{F}_{M}\mathbf{C}^{-1}_{n}\widehat{\mathbf{h}}^{\mathrm{c}}_{k_c})$
			\STATE \hspace{2em} Compute: $(\widetilde{\mathbf{y}}_{n})_{M\times 1}=\mathbf{F}_{M}\mathbf{C}_{n}^{-1}\mathbf{y}_{n}$
			\STATE \hspace{2em} Compute: $(\acute{\mathbf{y}}_{n})_{M\times 1}=\widetilde{\mathbf{y}}_{n}\oslash \mathrm{diag}^{-1}(\mathbf{D}_{n} \mathbf{D}_{n}^\mathrm{H} + \sigma^2 \mathbf{I}_M)$
			\STATE \hspace{2em} Compute: $\widehat{\mathbf{x}}_{(n-k_c)_N}=\mathbf{C}_{k_c}^{-1} \mathbf{C}_{n} \mathbf{F}_{M}^\mathrm{H} \mathbf{D}_{n}^\mathrm{H}\acute{\mathbf{y}}_{n}$
			\STATE \hspace{1em} \textbf{end for}
			\STATE \hspace{1em} \textbf{return}~$\widehat{\mathbf{X}}_{M \times N}=[\mathbf{\widehat{x}}_0, \mathbf{\widehat{x}}_1, \dots, \mathbf{\widehat{x}}_{N-1}]$
			
			\vspace{0.2em} 
			\STATE \textbf{Range reconstruction:}
			\STATE \hspace{1em} \textbf{Input:} estimated SAR channel gains $\widehat{\mathbf{H}}^\mathrm{r}$
			\STATE \hspace{1em} \textbf{Output:} reconstructed range profile ${\{\widehat{\beta}_q\}}_{q=0}^{Q-1}$
			\STATE \hspace{1em} \textbf{for}~$q=0$~\text{to}~$Q-1$~\textbf{do}
			\STATE \hspace{2em} Compute: $\widehat{\beta}_q \leftarrow q\text{-th element of}~\widehat{\mathbf{h}}^\mathrm{r}_{k_r}$
			\STATE \hspace{1em} \textbf{end for}
			\STATE \hspace{1em} \textbf{return}~${\{\widehat{\beta}_q\}}_{q=0}^{Q-1}$
		\end{algorithmic}
	\end{algorithm}
	
	\subsection{Performance Analysis}
	\begin{table*}[th]
		\centering
		\caption{Comparison of alternative channel sensing schemes}
		\label{tab1}
		\begin{tabular}{lccc}
			\toprule
			& \textbf{Complexity} & \textbf{Minimum PAPR} & \textbf{Minimum Noise Variance} \\
			\midrule
			PN Sequence [14] & $M^2N^2\log_2MN$ & $1$ & ${\sigma^2}/{E_0}$ \\
			CP-OTFS with Matched Filter [16] & $M^2N^2$ & $1$ & ${\sigma^2}/{E_0}$ \\
			Single Pilot Symbol [17] & $MN\log_2N$ & $M$ & ${\sigma^2}/{E_0}$ \\
			\textbf{Proposed Scheme} & $MN\log_2MN$ & $1$ & ${\sigma^2}/{E_0}$ \\
			\bottomrule
		\end{tabular}
	\end{table*}
	Since both the communication and SAR imaging processes rely on channel sensing, we first evaluate the performance of the proposed channel sensing scheme from the perspective of noise variance. Specifically, the noise variance is changed in the process of DD channel matrix extraction. We note that the noise vector ${\mathbf{z}}_{n}'$ in \eqref{eq22} can be expressed in terms of the AWGN vector ${\mathbf{z}}_{n}$ in \eqref{eq19} as
	\begin{equation} \label{eq31}
		{\mathbf{z}}_{n}'=\frac{1}{\sqrt{M}}{\mathbf{C}}_{n}{\mathbf{F}}^{\mathrm{H}}_{M}\left({\mathbf{F}}_{M}{\mathbf{C}}^{-1}_{n}{\mathbf{z}}_{n}\mathrm{\oslash }{\mathbf{F}}_{M}\mathbf{u}\right).
	\end{equation}
	Since the ${\mathrm{F}}_{M}$ and ${\mathbf{C}}_{n}$ operations are unitary, the noise variance of ${\mathbf{F}}_{M}{\mathbf{C}}^{-1}_{n}{\mathbf{z}}_{n}$ remains the same as ${\mathbf{z}}_{n}$.
	However, the element-wise division with ${\mathbf{F}}_{M}\mathbf{u}$ and $\mathbf{F}_{M}^\mathrm{H}$ operation change the noise variance of ${\mathbf{z}}_{n}'$ to $\sigma'^2=\frac{{\sigma }^{2}}{M^{2}}\sum^{M-1}_{m=0}{\left|{\tilde{u}}_m\right|}^{-2}$, where ${\tilde{u}}_m$ is the $m$-th element of ${\mathbf{F}}_{M}\mathbf{u}$. Notably, the noise variance of ${\mathbf{z}}_n'$ is minimized to $\frac{\sigma^2}{E_0}$ when all elements of ${\mathbf{F}}_M \mathbf{u}$ have the same energy, i.e., $\left|{\tilde{u}}_0\right|^2 = \cdots = \left|{\tilde{u}}_{M-1}\right|^2 = \frac{E_0}{M}$, where $E_0$ denotes the total energy of the pilot symbols. It is worth noting that the constant modulus property of the ZC sequence is preserved after the discrete Fourier transform \cite{cite28}. Thus, by adopting the ZC sequence as pilot symbols ${\mathbf{u}}$, the noise variance of ${\mathbf{z}}_n'$ is effectively minimized, while a low PAPR of the pilot signal is also achieved under the designed symbol pattern.
	
	Further, we analyze the performance of the proposed system architecture in both communication and SAR remote sensing. For communication, the MSE of the estimated CSI is defined as $\frac{{\left\lVert \widehat{\mathbf{H}}^\star- \widetilde{\mathbf{H}}^\star \right\rVert}^2}{M(K+1)}$, where $\widetilde{\mathbf{H}}^\star=\left[\mathbf{h}^\star_0,\dots,\mathbf{h}^\star_K\right]$, and is equal to the improved noise variance as
	\begin{equation} \label{eq32}
		\mathrm{MSE}=\frac{{\mathrm{\sigma }}^{2}}{E_0},
	\end{equation}
	which is equal to the noise variance of the alternative channel sensing schemes including PN-sequence-based matched filtering \cite{cite12}, CP-OTFS-based matched filtering \cite{cite13}, and SPA channel estimation \cite{cite15}. Notably, the MF-based schemes introduce additional interference that degrades the MSE performance, while the SPA-based signal structure inevitably results in a high PAPR, potentially affecting the MSE performance.
	Moreover, we propose a low-complexity MMSE method for channel equalization, which inherently ensures optimal MSE performance for the recovered data symbols.
	
	For the SAR remote sensing, the proposed scheme incorporates sufficient CP into the shared signals to achieve an IRCI-free range reconstruction. The reconstructed range profile is extracted from the estimated CSI obtained via the channel sensing scheme, allowing the signal-to-interference-plus-noise ratio (SINR) for the $q$-th range cell as follows
	\begin{equation} \label{eq33}
		\mathrm{SINR}_q = \frac{|{\beta}_q(\eta)|^2}{\frac{\sigma^2}{M^2} \sum_{m=0}^{M-1} |\tilde{u}_m|^{-2}},
	\end{equation}
	where the maximum SINR is achieved as \( |{\beta}_q(\eta)|^2 \frac{E_0}{\sigma^2} \) when \( |\tilde{u}_m|^2 = \frac{E_0}{M} \) for \( m \in \{0, 1, \dots, M-1\} \). Notably, the original SINR of the \( q \)-th range cell in the raw radar data is given as \( \frac{|{\beta}_q(\eta)|^2 E_0}{MN \sigma^2} \). Thus, after applying the proposed channel sensing scheme, the SINR performance improves by a factor of \( MN \), leading to a significant enhancement in the focusing of the radar data.
	
	Lastly, we analyze the computational complexity of the proposed integrated communication and remote sensing algorithm in terms of the required number of complex multiplications (CMs).
	We initially focus on the step of channel sensing, which is primarily governed by the operation in equation \eqref{eq22}, where the most computationally intensive step is the multiplication by $\mathbf{F}_{M}$. By employing the efficient $M$-point FFT and IFFT, this computation requires $\frac{M}{2} \log_2M$ CMs. Since equation \eqref{eq22} is performed $(K+1)$ times, the total computational complexity of channel sensing is $O(KM\log_2M)$.
	Subsequently, we consider the step of channel equalization, where the $N$ instances of $M$-point FFT operations yield a complexity of $O(MN\log_2M)$. Additionally, the ODDM demodulation process in \eqref{eq12} involves a computational load of $O(MN\log_2N)$.
	Combining these factors, the overall computational complexity of the proposed algorithm is $O(MN\log_2MN)$, consistent with that of standard OFDM communication systems, thus ensuring its practical feasibility.
	
	Notably, the proposed channel sensing scheme has a lower computational complexity than the state-of-the-art (SOTA) MF-based schemes \cite{cite13}, \cite{cite14}. Although the SPA scheme \cite{cite15} has the lowest computational complexity of $O(MN\log_2N)$, which is determined by the demodulation operation from the time domain to the DD domain, the high PAPR signal structure will inevitably degrading the performance of the power amplifier. The comparison of alternative DD-domain channel sensing schemes is summarized in Table \ref{tab1}.
	
	\section{Simulation and Prototype Validation}
	\label{sec5}
	In this section, we verify the effectiveness of the proposed protocol, architecture and algorithm by numerical simulation and prototype platform. The main simulation parameters for LEO satellite are set up according to 3GPP TR 38.821, which are summarized in Table \ref{tab2}.
	We configure the proposed integrated wireless frame to operate under the standard 5G NR configuration in sub-6 GHz scenarios, with $L_0 = 10 \, \text{ms}$, $F_0 = 15 \, \text{kHz}$, and $n_0 = 1$. Consequently, each time slot contains a constant $N_s=14$ number of CP-ODDM signals. Based on \eqref{eq16}, the lower boundary value of the numerology index is calculated as $\mu = 1$, which determines the subcarrier spacing as $\mathcal{F}=30\, \text{kHz}$. The scale of the ODDM signals is set to $M\mathrm{=}\mathrm{128}$ and $N\mathrm{=}\mathrm{32}$, resulting in an approximate ODDM signal bandwidth of $B_w=122.88\, \text{MHz}$.
	\begin{table}[th]
		\centering
		\caption{Main simulation parameters for LEO satellite-based integrated communication and remote sensing system}
		\label{tab2}
		\resizebox{0.92\linewidth}{!}{
			\begin{tabular}{@{\hspace{0.3cm}}l@{\hspace{0.5cm}} l@{\hspace{0.3cm}}}  
				\toprule
				\textbf{Parameter} & \textbf{Value} \\
				\midrule
				Orbit type & LEO \\
				Carrier frequency $f_0$ & 5 GHz \\
				Bandwidth $B_w$ & 122.88 MHz \\
				Orbit altitude $H$ & 550 km \\
				Satellite velocity $V_s$ & 7.6 km/s \\
				Doppler shift of satellite ${\upsilon}_{\scriptscriptstyle \mathrm{Sat}}$ & 63.4 kHz \\
				Squint angle $\theta_s$ & 30$^\circ$ \\
				Grazing angle $\theta_g$ & 69.5$^\circ$ \\
				Swath width $R_w$ & 1.2 km \\
				Aperture length $R_a$ & 3.6 km \\
				Satellite antenna gain $\omega$ & 30 dBi \\
				User antenna gain $G$ & 3 dBi \\
				Rain fading mean $\mu_r$ & -2.4 dB \\
				Rain fading variance $\sigma^2_r$ & 1.63 dB \\
				Rician factor $K_r$ & 5 \\
				Noise variance $\sigma^2$ & -106 dBm \\
				Power amplifier gain $P_0$ & 10 dBw \\
				\bottomrule
			\end{tabular}
		}
	\end{table}
	
	\subsection{Performance Simulation for Communication}
	We first simulate the communication performance of the proposed integrated algorithm. The data symbols are modulated using 4-quadrature amplitude modulation (QAM), and the transmitted signals are normalized to unit power, i.e., $E_0\mathrm{=}1$. Consequently, the power of the transmitted RF signal, denoted as $P_0$, equals to the gain of the power amplifier at the transmitter. For channel sensing, We utilize the ZC sequence as pilot symbols to ensure low PAPR of the pilot signal and minimize noise variance in channel sensing.
	
\begin{figure}[tbph]
		\centering
		\includegraphics[width=1\linewidth]{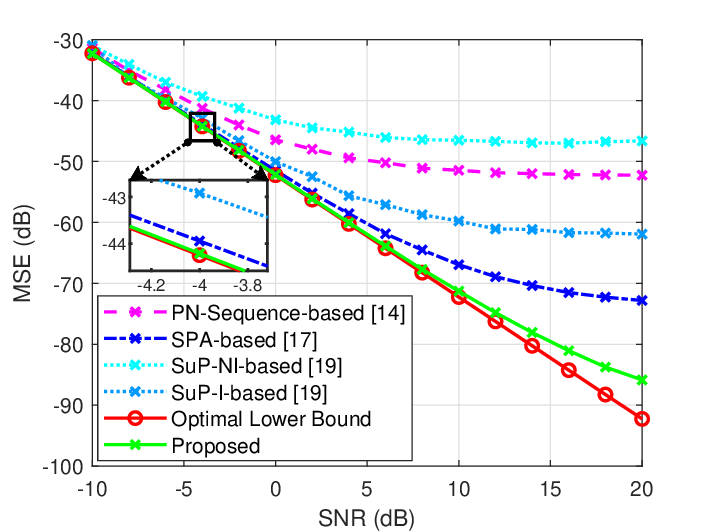}
		\captionsetup{labelformat=empty}
		\caption{\small Fig.~7.~~MSE comparison for DD-domain channel sensing.}
		\label{fig7}
	\end{figure}

	First, we compare the performance of the proposed channel sensing scheme with two baseline schemes across various signal-to-noise ratio (SNR) by adjusting the power amplifier gain $P_0$, where SNR is defined as $\mathrm{SNR}\mathrm{\ (dB)}\mathrm{=}\mathrm{10}{\mathrm{log}}_{\mathrm{10}}\mathrm{(}P_r/{\sigma }^2)$, where the received signal power $P_r=g^2P_0$. Among the three baseline schemes, the first is the PN-sequence-based scheme investigated in \cite{cite12}, which serves as a fundamental approach for channel prediction based on correlation detection.
	The second is the SPA scheme studied in \cite{cite15}, which has become one of the most widely adopted channel estimation techniques in OTFS-related research.
	Finally, the SOTA superimposed pilot-based method proposed in \cite{mishra2021otfs}, including both iterative (SuP-I) and non-iterative (SuP-NI) implementations, enhances spectral efficiency by superimposing data symbols and pilot signals. As demonstrated in Fig. \ref{fig7}, the proposed channel sensing scheme obtains the lowest MSE among the tested methods, owing to the low PAPR of the designed pilot signals and the significant SNR improvement provided by the proposed scheme. In comparison, the PN-sequence-based scheme suffers from interference when the code length is inadequate, which leads to suboptimal MSE performance. The SuP-based method also experiences mutual interference between data symbols and pilot signals, which can be progressively mitigated through iterative processing. Meanwhile, the SPA scheme results in a high PAPR for the transmitted signals, which adversely affects the performance of the power amplifier at the transmitter.
	
\begin{figure}[tbph]
		\centering
		\includegraphics[width=1\linewidth]{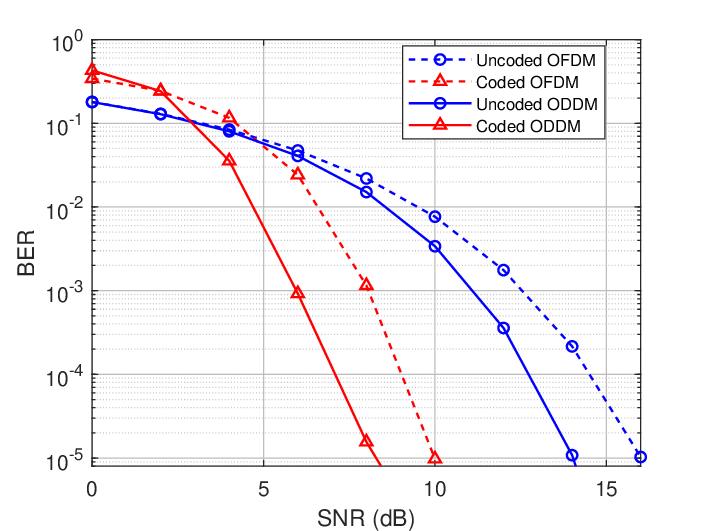}
		\captionsetup{labelformat=empty}
		\caption{\small Fig.~8.~~BER comparison for communication reception.}
		\label{fig8}
	\end{figure}

	We further simulate the BER performance of the communication receiver using the low-complexity MMSE channel equalization scheme, both with and without channel coding. For the coded systems, a rate 2/3 convolutional code with a constraint length of 7 and a puncture pattern of [1; 1; 0; 1] is employed, with the Viterbi decoding method used at the receiver. For comparison, we also evaluate the OFDM system with MMSE channel equalization.
	It can be seen clearly from Fig. \ref{fig8} that the uncoded ODDM outperforms the uncoded OFDM by about 2.1 dB at the BER of $1 \times 10^{-5}$. We also observe that about 1.8 dB performance
	gain is achieved for the coded ODDM over the coded
	OFDM at the BER of $1 \times 10^{-5}$.
	Notably, the proposed scheme maintains a low computational complexity similar to that of standard OFDM, while achieving notable BER gains. This highlights both its performance advantages and practical suitability under LEO satellite channel conditions.
	Since BER improves as SNR increases, the communication quality of LEO satellite systems can be enhanced by achieving higher gain from the power amplifier at the satellite transmitter.
	
	\subsection{Performance Simulation for SAR Imaging}
	Herein, we present simulation results for the performance of SAR imaging. The IRCI-free range reconstruction is effectively achieved through unified DD-domain channel sensing. Subsequently, the azimuth processing adopts the conventional approach for stripmap SAR imaging \cite{cite26}, including RCMC and azimuth compression.
	
\begin{figure}[htbp]
		\centering
		\begin{minipage}{0.9\linewidth}
			\centering
			\includegraphics[width=1\linewidth]{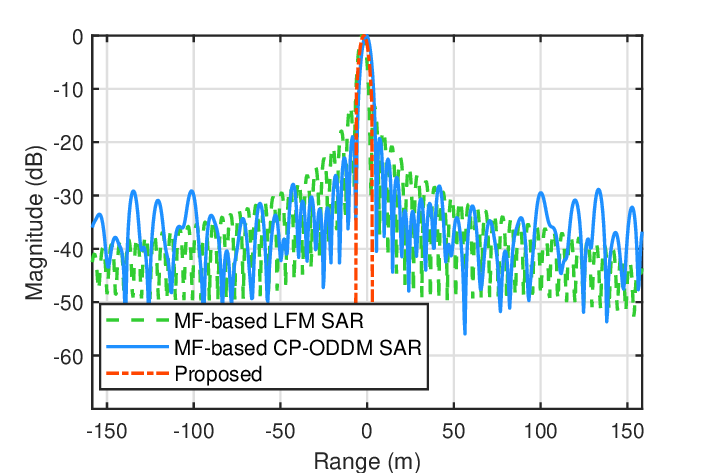}
			\captionsetup{labelformat=empty,justification=centerfirst}
			\caption*{\small (a)~Range Profile}
			\label{fig9.1}
		\end{minipage}%
		\hfill
		\begin{minipage}{0.9\linewidth}
			\centering
			\includegraphics[width=1\linewidth]{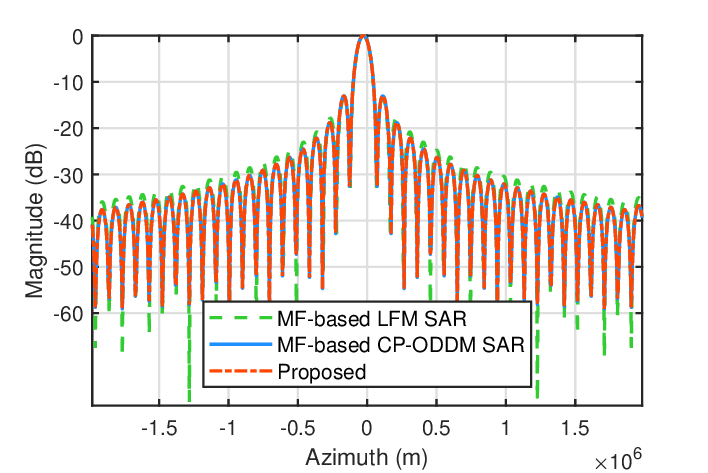}
			\captionsetup{labelformat=empty,justification=centerfirst}
			\caption*{\small (b)~Azimuth Profile}
			\label{fig9.2}
		\end{minipage}
		\captionsetup{labelformat=empty}
		\caption{\small Fig.~9.~~Point spread function comparison.}
		\label{fig9}
	\end{figure}

	Fig. \ref{fig9} demonstrates the normalized range profiles and azimuth profiles of the point spread function. For comparison, we also evaluate the LFM signals and random-coded CP-ODDM signals using the conventional range Doppler algorithm (RDA) for MF-based pulse compression \cite{cite26}. It can be observed that the proposed scheme achieves IRCI-free range reconstruction, whereas the LFM and MF-based CP-ODDM SAR imaging schemes exhibit high range side-lobes, which interfere with adjacent range cells. Furthermore, the azimuth profiles of the point spread function are similar for these three SAR imaging schemes.
	
\begin{figure}[htbp]
		\centering
		\begin{minipage}{0.49\linewidth}
			\centering
			\includegraphics[width=1\linewidth]{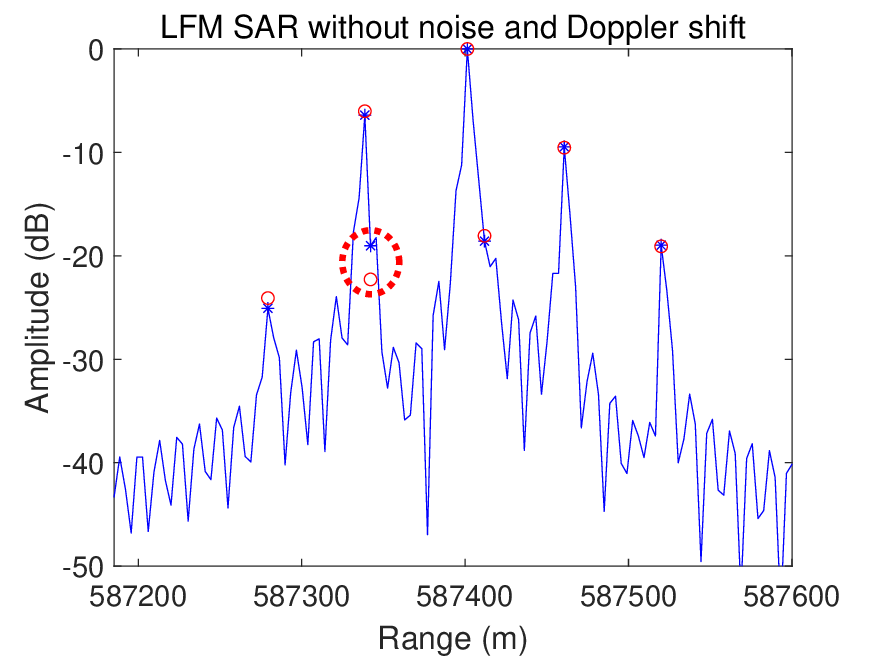}
			\captionsetup{labelformat=empty,justification=centerfirst}
			\caption*{\small (a)}
			\label{fig10.1}
		\end{minipage}%
		\hfill
		\begin{minipage}{0.49\linewidth}
			\centering
			\includegraphics[width=1\linewidth]{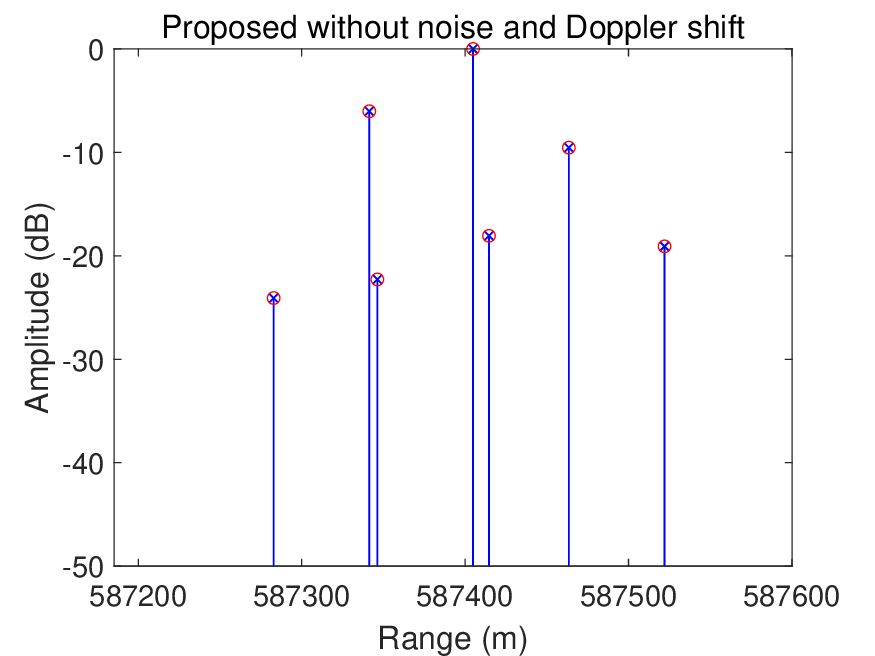}
			\captionsetup{labelformat=empty,justification=centerfirst}
			\caption*{\small (b)}
			\label{fig10.2}
		\end{minipage}
		\begin{minipage}{0.49\linewidth}
			\centering
			\includegraphics[width=1\linewidth]{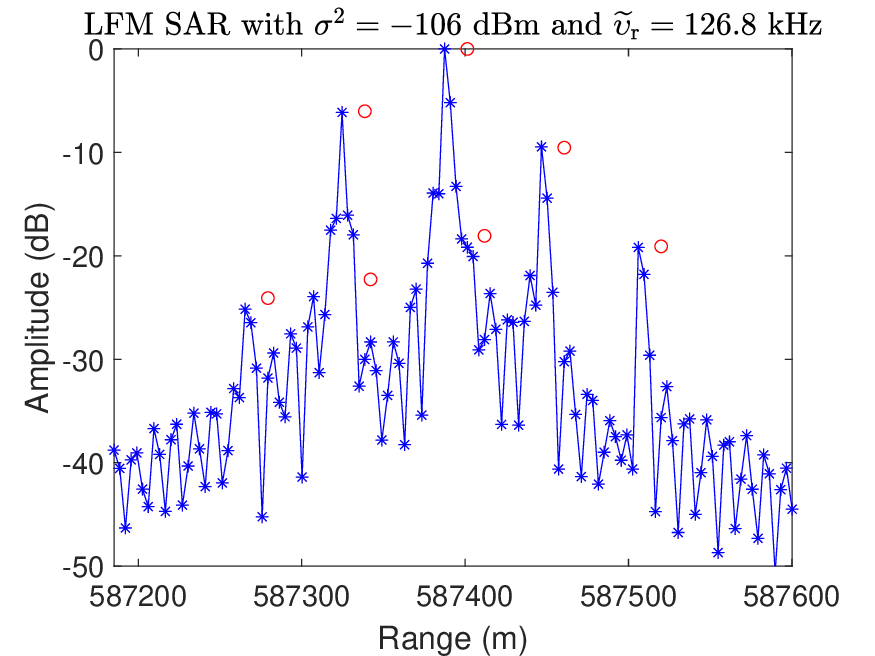}
			\captionsetup{labelformat=empty,justification=centerfirst}
			\caption*{\small (c)}
			\label{fig10.3}
		\end{minipage}
		\begin{minipage}{0.49\linewidth}
			\centering
			\includegraphics[width=1\linewidth]{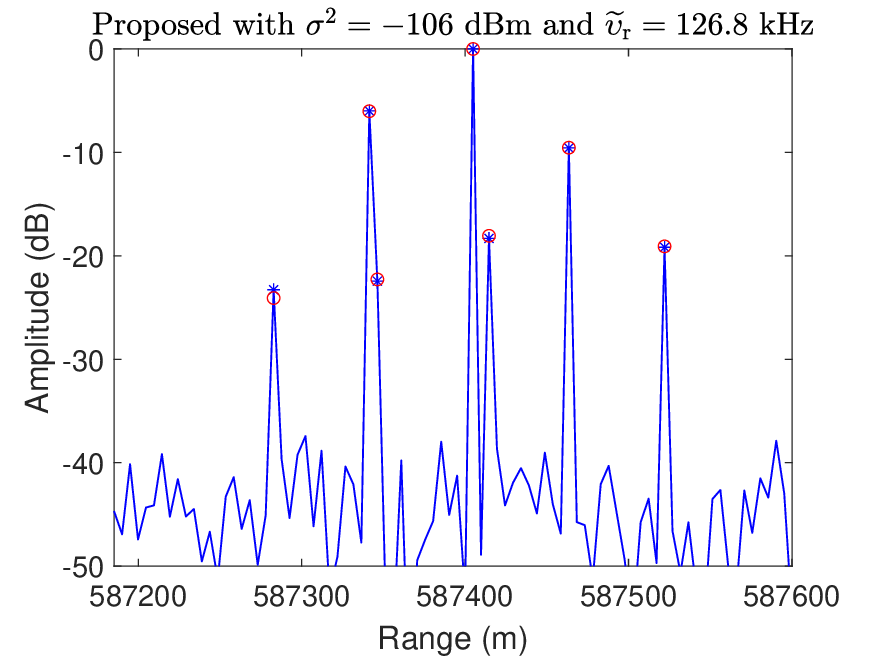}
			\captionsetup{labelformat=empty,justification=centerfirst}
			\caption*{\small (d)}
			\label{fig10.4}
		\end{minipage}
		\captionsetup{labelformat=empty}
		\caption{\small Fig.~10.~~A range line of imaging results. Red circles denote the real RCS coefficients of targets, blue asterisks represent the reconstructed range profiles.}
		\label{fig10}
	\end{figure}

	We also consider a single range line with $Q = 128$ range cells in a $R_w = 1200 \, \text{m}$ wide beam swath, and targets (non-zero RCS coefficients) are included in 7 range cells located from $587 \, 282 \, \text{m}$ to $587 \, 522 \, \text{m}$, the amplitudes are randomly generated, shown as the red circles in Fig. \ref{fig10}. For comparison, the transmitted LFM pulse duration and bandwidth are set to be the same as the shared ODDM signals. The normalized range line imaging results are shown as the blue asterisks in Fig. \ref{fig10}. The results demonstrate that the range profile of the proposed SAR imaging scheme aligns precisely with the normalized RCS coefficients in the absence of noise and intra-pulse Doppler shift. However, the range side-lobes in LFM SAR cause interference from nearby targets, which can obscure weaker targets and prevent accurate imaging. Further considering the noise in raw radar data with variance ${\sigma }^2=-106\ \mathrm{dBm}$ and the intra-pulse Doppler shift $\widetilde{\upsilon }_r=\frac{\mathrm{2}V_sf_0}{c}\mathrm{sin}\mathrm{(}{\theta }_s\mathrm{)}\mathrm{=}\mathrm{126}.\mathrm{8\ kHz}$, the simulation results indicate that the proposed SAR imaging scheme effectively mitigates the impacts of intra-pulse Doppler shift and noise, maintaining the accuracy of SAR imaging. In contrast, the range profile in LFM SAR suffers from significant distortion due to the variations in the ambiguity function under different Doppler shifts. Thus, the performance advantage of the proposed SAR imaging scheme over LFM SAR is evident due to its IRCI-free range reconstruction and its ability to counteract Doppler shifts.
	
	\subsection{Prototype Verification}
	To verify the effectiveness of the proposed integrated communication and remote sensing system for real-time SAR imaging and information transmission, we develop an SDR-based prototype platform that implements the integrated system architecture, protocol, and algorithm outlined in this paper.
	\subsubsection{Hardware Composition}
	 The equipment of the SDR-based prototype platform includes universal software radio peripheral (USRP), mmWave phased array (mmPSA), mmWave converter, unmanned aerial vehicle (UAV), etc. The specifications of the main hardware components supporting signal transmission in the mmWave band are as follows.
	\begin{itemize}
		\item
		\textbf{USRP}: The NI USRP-X410 is a high-performance SDR platform designed by National Instruments. It offers a real-time bandwidth of up to 400 MHz and is equipped with four daughter channels covering a frequency range of 1 MHz to 7.2 GHz.
		\item
		\textbf{mmPSA}: The mmPSA-TR64A is a $16$-element mmWave phased array with each antenna element corresponding to a 4-element vertical sub-array, enabling individual control of each element for effective beamforming. The device is suitable for mmWave communication, enabling the transmission of 28 GHz Ka-band signals.
		\item
		\textbf{mmWave Converter}: The UD Box 5G mmWave converter is a frequency converter that supports up- and down-conversion between intermediate frequency (IF) signals in the range of 0.01-14 GHz and RF signals in the 24-44 GHz range.
	\end{itemize}
	
	\begin{figure}[tbph!]
		\centering
		\includegraphics[width=0.92\linewidth]{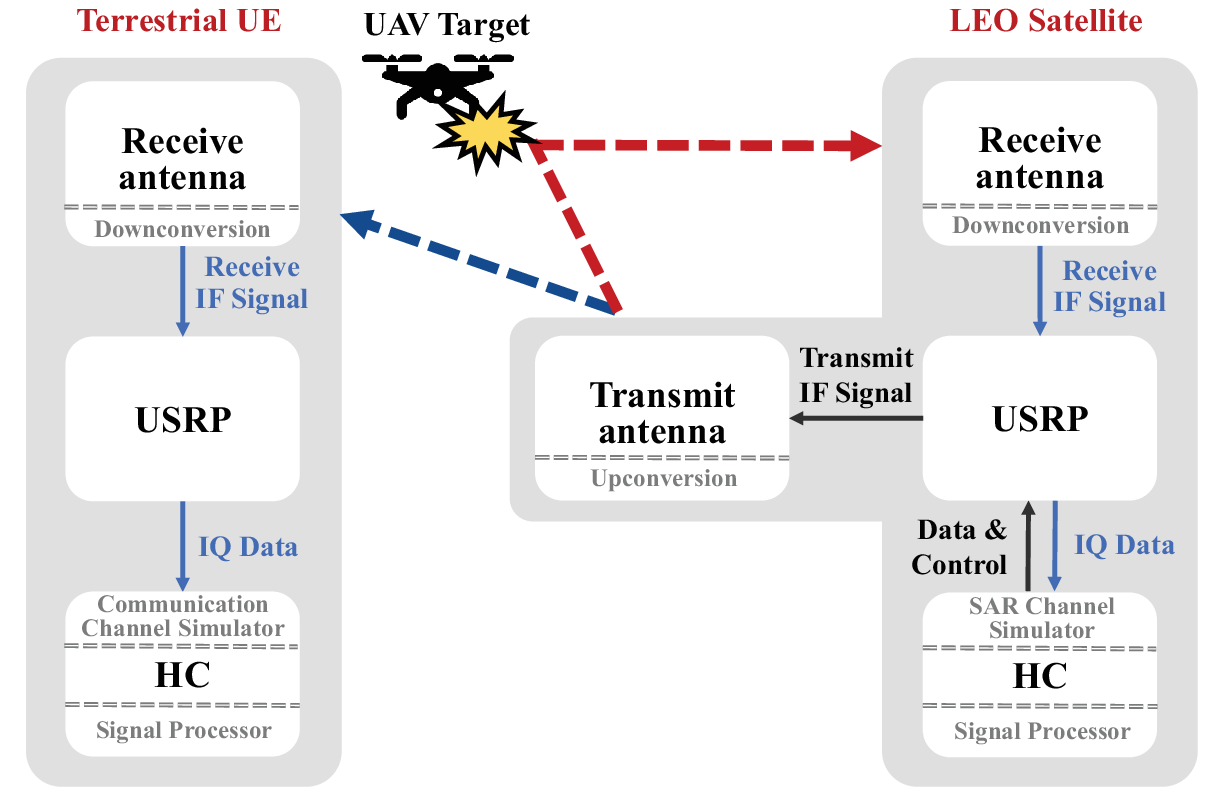}
		\captionsetup{labelformat=empty}
		\caption{\small Fig.~11.~~Program design of the prototype platform.}
		\label{fig11}
	\end{figure}
	
	As illustrated in Fig. \ref{fig11}, the prototype platform primarily consists of three components: a LEO satellite, a terrestrial UE, and a UAV target. The satellite consists of a host computer (HC) connected to a dual-channel USRP, with two mmPSAs interfacing via mmWave converters to function as a versatile RF front-end. The host computer implements the proposed transmission protocol and system architecture, as shown in Fig. \ref{fig4} and \ref{fig6}, output IQ data to USRP for transferring and also simulates the large Doppler shift and multipath propagation characteristics of SAR channels for received IQ data. Furthermore, the transmit channel of the satellite USRP outputs IF signal at $2.8 \, \text{GHz}$, which is then upconverted and transmitted through the transmitting mmPSA at $28 \, \text{GHz}$. Consistent with the system model, the transmitted beam concurrently covers the receiving antenna of the UE and the UAV target.
	The echo signal from the UAV is captured by the satellite’s receiving mmPSA, downconverted, and processed for SAR imaging.
	As for the terrestrial UE component, another single-channel USRP equipped with a mmPSA and a mmWave converter is used for communication reception, forwarding the processed IQ signals to another HC, which simulates the downlink channel characteristics and performs communication processing.
	
	As described above, the prototype platform simultaneously enables downlink communication and SAR remote sensing functionalities, consistent with the core capabilities of the proposed integrated communication and remote sensing system.
	\begin{table}[ht]
		\centering
		\caption{Main experimental configurations of the prototype system}
		\label{tab3}
		\small
		\resizebox{1\linewidth}{!}{
			\begin{tabular}{@{\hspace{0.2cm}}l@{\hspace{0.2cm}}l@{\hspace{0.4cm}}l@{\hspace{0.2cm}}l@{\hspace{0.2cm}}}
				\toprule
				\textbf{Parameter} & \textbf{Value} & \textbf{Parameter} & \textbf{Value} \\
				\midrule
				Carrier frequency & 28 GHz & Antenna Array & $8\times 8$ \\
				Modulation alphabet & 16-QAM & Numerology index & 4 \\
				IQ rate & 491.52 MHz & Subcarrier spacing & 120 kHz \\
				Bandwidth & 245.76 MHz & Transmission rate & 561.74 Mbps \\
				Range resolution & 0.61 m & Azimuth resolution & 0.77 m \\
				\bottomrule
			\end{tabular}
		}
	\end{table}
	
	\subsubsection{Experimental Results}
	Herein, we present the experimental configurations of the prototype system at carrier frequencies of $28 \, \text{GHz}$, as depicted in Table \ref{tab3}. Compared to the sub-6 GHz scenarios, the reference frequency constant is reset to $F_0 = 7.5 \, \text{kHz}$, the input DD-domain symbols are modulated using 16-QAM, and the scale of ODDM signals is set to $M=128$ and $N=16$, while other parameters remain unchanged.
	As a result, each time slot contains $N_s = 7$ CP-ODDM signals. The numerology index is calculated as $\mu = 4$, which determines the subcarrier spacing as $\mathcal{F} = 120 \, \text{kHz}$, yielding an approximate ODDM signal bandwidth of $B_w = 245.76 \, \text{MHz}$. Additionally, a two-time oversampling rate is set for DDOP-based pulse shaping and matched filtering, with an IQ rate of $491.52 \, \text{MHz}$.
	\begin{figure*}[tbph!]
		\centering
		\includegraphics[width=1\linewidth]{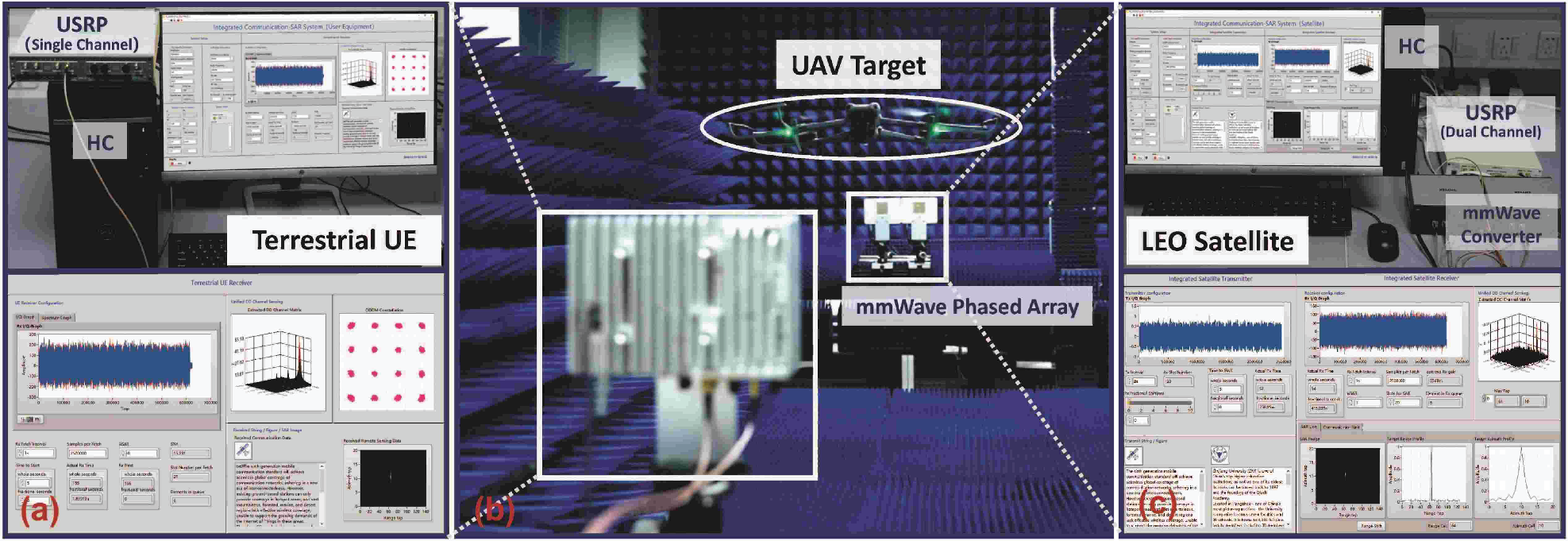}
		\captionsetup{labelformat=empty}
		\caption{\small Fig.~12.~~Real scenario of the prototype platform.}
		\label{fig12}
	\end{figure*}
	
	The real experimental scenario is depicted in Fig. \ref{fig12}. Specifically, in our experiment workflow, the LEO satellite initially sends a message and an image as communication data to UE within a $10 \, \text{ms}$ wireless frame consisting of $80$ time slots. The bit rate of data transmission is $561.74 \, \text{Mbps}$, while the range and azimuth resolutions of SAR imaging are $\rho_r=0.61\, \text{m}$ and $\rho_a=0.77\, \text{m}$, respectively. Since the scenario is adorned with the mmWave-absorbing materials, the UAV target can be seen the only scattering object located in a single range cell of the beam swath. When we adjust the propagation range between the UAV and the satellite, the range cell will migrate accordingly.
	It can be seen clearly from Fig. \ref{fig12}(c) that the reconstructed range profile obtained using the proposed scheme remains low-sidelobe dispersion as a result of DDOP-based pulse shaping and matched filtering, thus validating the IRCI-free imaging characteristics of the proposed SAR imaging scheme with significant intra-pulse Doppler shift.
	It is worth noting that the resulting SAR image will also be encoded and modulated by the satellite, then transmitted to the UE along with other communication data in the subsequent wireless frame.
	
	For terrestrial UE, the received signal undergoes real-world hardware imperfections, synchronization errors, and simulated satellite-to-ground channel conditions specified in Table \ref{tab2}, thereby effectively capturing the main signal characteristics of a real LEO satellite downlink. According to the proposed processing architecture, the UE first detects the wireless frame transmitted by the satellite, mitigates the effects of fractional delays using DDOP-based matched filtering, and compensates for Doppler shifts using CP-based CFO elimination \cite{cite27}. Then, the proposed channel sensing and equalization process is applied to mitigate the on-grid DD spreading effects. As demonstrated in Fig.~\ref{fig12}(a), even though the extracted DD channel matrix exhibits significant off-grid dispersion, the accurate post-equalization constellation points confirm that the proposed processing architecture effectively mitigates the challenges encountered in a real-world propagation and accurately recovers the communication and remote sensing data from the satellite.
	
	The above experimental results effectively validate the real-time SAR imaging and information transmission capabilities of the proposed integrated communication and remote sensing system.
	
	\section{Conclusion}
	\label{sec6}
	In this paper, we designed and implemented the integration of communication and SAR remote sensing in LEO satellite systems. We first provided an ODDM-based system architecture for real-time SAR imaging and information transmission in satellite direct-connect UE scenarios.
	Then, we specifically proposed an efficient transmission protocol, including a wireless frame structure based on 5G NR standard and a low-PAPR pilot signal pattern.
	Next, we designed a unified signal processing framework for the satellite receiver that simultaneously supports information communication against the DD spreading channels and IRCI-free range reconstruction for SAR imaging.
	Finally, we performed extensive simulations and constructed a prototype platform to validate the feasibility of the proposed integrated communication and remote sensing system for real-time SAR imaging and information transmission.
	
	\bibliographystyle{IEEEtran}
	\bibliography{IEEEabrv,ref}

\end{document}